\documentclass[sigplan, screen]{acmart}
\usepackage{amsmath,amsfonts}
\usepackage{algorithm}
\usepackage{algpseudocode}
\usepackage{multicol}
\usepackage{paracol}
\usepackage{geometry}
\usepackage{graphicx}
\usepackage{wrapfig}
\usepackage{pgfplots}
\usepackage{caption}
\usepackage{subcaption}
\usepackage{fancyhdr}
\usepackage[parfill]{parskip}
\usepackage{hyperref}
\usepackage{amsmath}
\usepackage{listings}
\usepackage{mathtools}
\usepackage{array}
\usepackage{float}
\usepackage{booktabs}
\usepackage{enumitem}
\usepackage{wrapfig}
\usepackage{svg}
\usepackage[edges]{forest}
\usepackage{multirow}
\usepackage{accents}
\usepackage{xspace}
\usepackage{tabularx}
\usepackage[title]{appendix}

\copyrightyear{2025}
\acmYear{2025}
\setcopyright{acmlicensed}
\acmConference[PPoPP '25]{The 30th ACM SIGPLAN Annual Symposium on Principles and Practice of Parallel Programming}{March 1--5, 2025}{Las Vegas, NV, USA}
\acmBooktitle{The 30th ACM SIGPLAN Annual Symposium on Principles and Practice of Parallel Programming (PPoPP '25), March 1--5, 2025, Las Vegas, NV, USA}
\acmDOI{10.1145/3710848.3710887}
\acmISBN{979-8-4007-1443-6/25/03}

\newcommand{\R}{\mathbb{R}}

\newcommand{\cO}{\mathcal{O}}
\newcommand{\diag}{\texttt{diag}}
\newcommand{\V}{$\mathbf{V}$\xspace}
\newcommand{\Pmat}{$\mathbf{P}$\xspace}
\newcommand{\C}{$\mathbf{C}$\xspace}
\newcommand{\K}{$\mathbf{K}$\xspace}

\newcommand{\B}{$\mathbf{B}$\xspace}

\newcommand{\D}{$\mathbf{D}$\xspace}
\newcommand{\z}{$\mathbf{z}$\xspace}

\definecolor{pblue}{rgb}{0.13,0.13,1}
\definecolor{pgreen}{rgb}{0,0.5,0}
\definecolor{pred}{rgb}{0.9,0,0}
\definecolor{pgrey}{rgb}{0.46,0.45,0.48}

\usepackage[most]{tcolorbox}

\tcbset{width=\columnwidth,boxrule=0pt,colback=red,arc=0pt,auto outer arc,left=0pt,right=0pt,boxsep=5pt}

\definecolor{customcolor}{HTML}{9494b8}

\newtcolorbox[auto counter]{ObservationBox}{
    borderline west={3pt}{0pt}{DeepSkyBlue4},
    colback=customcolor!30!white}
\newtcolorbox{ContributionBox}{textmarker,
    borderline west={3pt}{0pt}{green},
    colback=green!10!white}
\newtcolorbox{LimitationBox}{textmarker,
    borderline west={3pt}{0pt}{red},
    colback=red!10!white}

\def\BibTeX{{\rm B\kern-.05em{\sc i\kern-.025em b}\kern-.08em
    T\kern-.1667em\lower.7ex\hbox{E}\kern-.125emX}}

\usepackage{multirow}
\usepackage{tabularx}
\newcolumntype{e}{>{\scriptsize}p}

\DeclareMathOperator*{\argmin}{arg\,min}

\def\nullvalue(#1){#1}
\def\drawarraystack(#1, #2){
  \begin{tikzpicture}
    \def\scale{0.6}
    \def\array{#1}
    \foreach \t/\i/\v/\n in \array {
      \node[minimum size=\scale cm] at (\i * \scale, \scale - 0.1) (index\i) {\footnotesize \t};
      \ifthenelse{\equal{\n}{n}}{
        \def\fillcolor{lightgray}
      }{
        \def\fillcolor{lightgray!40}
      }
      \node[draw, fill=\fillcolor, minimum size=\scale cm] at (\i * \scale, 0) {\scriptsize\v};
    }
    \node at (-1, 0) (start) {#2};
    \node at (-1, 0.5) (start) {\footnotesize Thread idx};
  \end{tikzpicture}
}

\usepackage{xspace}
\newcommand{\kmeans}{K-means\xspace}
\newcommand{\kmeanspp}{K-means++\xspace}
\newcommand{\kk}{Kernel~K-means\xspace}
\newcommand{\popcorn}{\textsc{Popcorn}\xspace}

\pgfplotsset{compat=1.18}

\begin{document}

\title{Popcorn: Accelerating Kernel \kmeans on GPUs through Sparse Linear Algebra} 
\author{Julian Bellavita\texorpdfstring{$^{*}$}{*}}\thanks{*Corresponding authors\newline*Part of this work was done while the first author was a visiting student at the University of Trento.}
\affiliation{
    \institution{Cornell University}
    \city{Ithaca, NY}
    \country{USA}
}
\email{jbellavita@cs.cornell.edu}

\author{Thomas Pasquali}
\affiliation{
    \institution{University of Trento}
    \city{Trento}
    \country{Italy}
}
\email{thomas.pasquali@unitn.it}

\author{Laura Del Rio Martin}
\affiliation{
    \institution{University of Trento}
    \city{Trento}
    \country{Italy}
}
\email{laura.delrio@unitn.it}

\author{Flavio Vella\texorpdfstring{$^{*}$}{*}}
\affiliation{
    \institution{University of Trento}
    \city{Trento}
    \country{Italy}
}
\email{flavio.vella@unitn.it}

\author{Giulia Guidi\texorpdfstring{$^{*}$}{*}}
\affiliation{
    \institution{Cornell University}
    \city{Ithaca, NY}
    \country{USA}
}
\email{gguidi@cs.cornell.edu}

\begin{abstract}

\kmeans is a popular clustering algorithm with significant applications in numerous scientific and engineering areas.
One drawback of \kmeans is its inability to identify non-linearly separable clusters, which may lead to inaccurate solutions in certain cases. 
\kk is a variant of classical \kmeans that can find non-linearly separable clusters.
However, it scales quadratically with respect to the size of the dataset, taking several minutes to cluster even medium-sized datasets on traditional CPU-based machines.

In this paper, we present a formulation of \kk using sparse-dense matrix multiplication (SpMM) and sparse matrix-vector multiplication (SpMV), and we show that our formulation enables the rapid implementation of a fast GPU-based version of \kk with little programming effort.
Our implementation, named \popcorn, is the first open-source GPU-based implementation of \kk.

\popcorn achieves a speedup of up to 123.8$\times$ over a CPU implementation of \kk and a speedup of up to 2.6$\times$ over a GPU implementation of \kk that does not use sparse matrix computations.
Our results support the effectiveness of sparse matrices as tools for efficient parallel programming.

\end{abstract}



\maketitle

\section{Introduction}

 \kmeans clustering is one of the most popular and powerful data mining algorithms.
 Its applications are numerous, ranging from economics~\cite{noviandy2024environmental,wielechowski2021interdependence}, computational biology~\cite{melman2018k}, approximate matrix factorization~\cite{chen2022randomly}, and vector quantization~\cite{blalock2021multiplying}.
Although it has many advantages, a key limitation of \kmeans is that it can only find linearly separable clusters.
This can cause \kmeans to compute inaccurate clusterings for certain datasets.
\kk is a variant of \kmeans that can find non-linearly separable clusters~\cite{dhillon2004kernel}, and therefore does not suffer from this limitation.
As with classical \kmeans, \kk is used in a variety of areas ranging from computational biology~\cite{appCancer, appDiabetes, appDiabetes1}, performance prediction~\cite{appStudents}, smart agriculture~\cite{appAgricolture} and more.


\kk projects the data points from their original input space into a high-dimensional feature space and clusters them in this high-dimensional feature space.
Cluster boundaries that are linear in the high-dimensional feature space can be non-linear in the original input space, meaning \kk can find non-linearly separable clusters.
Despite its advantages over classical \kmeans, \kk scales poorly with respect to the size of the data set.
If $n$ is the number of points in the dataset and $d$ is the number of features in the dataset, each iteration of \kk requires $\cO(n^2)$ computation.
In addition, there is a preprocessing step that requires $\cO(n^2d)$ computation.
This high computational cost means that runs of \kk on traditional CPU-based architectures take several minutes, even for medium-sized datasets.
In particular, for latency-sensitive or real-time applications, such as image change detection~\cite{imageChange}, these runtimes are not acceptable.

One potential option for reducing the long runtimes of \kk is GPU acceleration. 
A GPU-capable version of \kk is particularly appropriate because the preprocessing step and the clustering itself both involve significant amounts of computation.
The superiority of GPUs over CPUs for these sorts of compute-intensive applications has been well established over the last 15 years~\cite{GPUdl, GPUSC, vestias2014trends}. 
Furthermore, GPU-based implementations of classical \kmeans have been shown to achieve significant speedups \cite{martin2020gpukmeans}, suggesting that a GPU-based version of \kk would achieve similar outcomes.
To the best of our knowledge, there is no open-source GPU implementation of \kk.
A GPU implementation of \kk has been described in the literature~\cite{baydoun2018cpu}, but it is not publicly available.
Other existing GPU implementations, such as RAPIDS~\cite{rapidsai} and the work of Markovtsev et al.~\cite{vadim_markovtsev_2017_286944}, provide support for classical \kmeans, but not for \kk.

GPUs can provide significant speedup, but they are difficult to program to maximize performance.
Programming GPUs using a framework such as CUDA~\cite{nickolls2008scalable} involves considerable effort and manual tuning to achieve good performance, and the programmer must reason about the low-level GPU behavior
\cite{Sato2010, cudaHard}.


In this paper, we present a strategy for implementing \kk on GPUs based on sparse linear algebra that requires minimal manual programming effort while still achieving high performance.
First, we introduce a formulation of \kk that casts most of the computation in terms of sparse-dense matrix multiplication (SpMM) and sparse matrix-vector multiplication (SpMV).
Then, we describe a GPU implementation based on this formulation that uses the cuSPARSE~\cite{cuSPARSE} and cuBLAS~\cite{cuBLAS} libraries to perform most of the computation. 
Finally, we show that our formulation of \kk using SpMM and SpMV simplifies the implementation of a performant GPU version of \kk.
Our implementation is named \popcorn, and it is the first open-source GPU version of \kk.

\popcorn achieves a speedup of up to 123.8$\times$ compared to the fastest CPU implementation of \kk on several real-world datasets from different scientific and engineering areas and a speedup of up to $2.6\times$ compared to a baseline CUDA implementation.
Our implementation is publicly available at \url{https://github.com/HicrestLaboratory/Matrix-Centric-K-Means}
Our results demonstrate the viability of sparse matrix computations as tools for productive and performant parallel programming on GPUs without significant kernel engineering effort.

\noindent
Our main contributions include:

\begin{enumerate}
    \item A new formulation of \kk in terms of the sparse linear algebra primitives SpMM and SpMV; 
    \item A strategy for dynamically selecting the best matrix-centric algorithm to compute the kernel matrix based on the number of points and the number of features in the dataset;
    \item \popcorn, the first open-source GPU-based implementation of \kk;
    \item \popcorn achieves up to a $2.6\times$ speedup over a GPU-based implementation of \kk that does not use sparse matrices and a $123.8\times$ speedup over a CPU-based implementation.
\end{enumerate}


\section{Background}\label{sec:background}

In this section, we provide background information on classical \kmeans and \kk, emphasizing the differences between the two approaches.

\begin{table}[t]
\centering
\caption{A list of important symbols used in the paper.}
\label{tab:symbols}
\begin{tabularx}{\columnwidth}{ll}
    \hline
    Symbol & Description \\
    \hline
    $d$ & number of features \\
    $n$ & number of points \\
    $k$ & number of clusters, $k < n$ \\
    $c_k$ & centroid of $L_k$, $c_k\in \mathbb{R}^d$\\
    \D & distances matrix \\
    \K & kernel matrix \\
    \V & sparse selection matrix \\
    \hline
\end{tabularx}
\end{table}

\subsection{Classical \kmeans}

\kmeans is an optimization problem that partitions a set of points $P = \{ p_1, \hdots, p_n\}$ into $k$ clusters $L_1,...,L_k$ with centroids $c_1,...,c_k$ such that the sum of the squared Euclidean distance between each point and its nearest centroid is minimized~\cite{lloyd}.
Formally, the \kmeans objective function is:

\begin{equation*}
    \min\left (\sum\limits^k_{j=1}\sum\limits_{p_i\in L_j} \left\|p_i-c_j\right\|^2\right)
\end{equation*}

\noindent
Finding a globally optimal solution to a \kmeans problem is NP-hard, but there are effective heuristic algorithms that can find local optima.
The canonical \kmeans algorithm of this type is known as Lloyd's algorithm~\cite{lloyd}.
Lloyd's algorithm has three phases:

\begin{enumerate}
    \item Calculate the distance between each point and centroid;
    \item Determine the closest centroid to each point and assign each point to the corresponding cluster;
    \item Calculate new centroid positions using a weighted mean of the points assigned to each cluster.
\end{enumerate}

\noindent
Theses phases are repeated until convergence or until a set number of iterations have been executed. 
The time complexity of each iteration of Lloyd's algorithm is $\mathcal{O}(ndk)$.

Lloyd's algorithm converges to a local optimum, but the quality of the optimum is strongly dependent on the chosen initial centroid locations.
A simple strategy for choosing initial centroid locations is to sample $k$ points uniformly at random and set the initial centroid locations to be the $k$ sampled points.
This method is inexpensive but can cause \kmeans to converge to a poor local optimum.
An alternative approach that is specifically designed to find good local optima is the \kmeanspp initialization algorithm \cite{kmeans++}.
\kmeanspp chooses the initial centroids by randomly sampling points according to a weighted probability distribution.
This ensures that points that are far away from previously selected points are more likely to be selected as centroids in subsequent iterations.
\kmeanspp is computationally expensive, but it guarantees that a solution within a $\cO(\log k)$ factor of the optimal solution is found.


\subsection{\kk}

Classical \kmeans is unable to identify non-linearly separable clusters, which can lead to inaccurate solutions for certain datasets~\cite{dhillon2004kernel, kernelmethods}.
To overcome this limitation, the \kk algorithm was proposed~\cite{dhillon2004kernel}.
\kk projects the points of the dataset from their original input space into a high-dimensional feature space before they are clustered.
Clustering in the high-dimensional feature space can lead to cluster boundaries that are nonlinear in the original input space, resulting in nonlinear clusters and, in some cases, more accurate clusterings.
\kk not only provides better clusterings than classical \kmeans in some cases but it has also been shown to be equivalent to spectral clustering~\cite{dhillon2004kernel}, another popular clustering algorithm.

Clustering in the feature space requires computing the pairwise inner products of the points.
Instead of calculating the coordinates of points in the high-dimensional feature space and then determining their inner product, it is possible to calculate the inner product while leaving the points in their original input space using a technique called the ``kernel trick''~\cite{kernelmethods}.
The kernel trick involves applying a non-linear kernel function $\kappa(\mathbf{x, y})$ to the vectors $\mathbf{x}$ and $\mathbf{y}$. 
If $\kappa(\mathbf{x, y})$ is chosen appropriately, it computes the inner product of $\mathbf{x}$ and $\mathbf{y}$ in a high-dimensional feature space without needing to project either vector into that feature space.
Typically, $\kappa(\mathbf{x, y})$ is evaluated for each pair of points in the dataset, and the results are stored in a \textit{kernel matrix} \K:

\[
\mathbf{K} = \begin{bmatrix}
\kappa(p_1, p_1) & \hdots & \kappa(p_1, p_n) \\
\vdots & \ddots & \vdots \\
\kappa(p_n, p_1) & \hdots & \kappa(p_n, p_n)
\end{bmatrix}
\]



\begin{algorithm}[t]
\caption{\kk Algorithm}
\label{alg:kkmeans}
\begin{algorithmic}
\Require Set of points $P=\{p_1,...,p_n\}$ where $p_i\in\mathbb{R}^d$, an integer $k < n$, maximum iterations $maxiter$, non-linear projection function $\phi$
\State Initialize iteration counter $iter = 0$
\While{$iter < maxiter$ and centroids are changing}
    \ForAll{$p_i\in P$}
        \State $j \gets \argmin_j\left\|\phi(p_i)-c_j\right\|^2$ using the kernel trick.
        \State Add $p_i$ to cluster $L_j$
    \EndFor
    \State $iter \gets iter + 1$
\EndWhile
\end{algorithmic}
\end{algorithm}

\kk is more computationally expensive than classical \kmeans.
Unlike with classical \kmeans, \kk requires computing the kernel matrix \K.
This requires $\cO(n^2d)$ work, which is significant compared to the $\cO(ndk)$ cost of a single iteration of Lloyd's algorithm, given that $k$ is almost always much smaller than $n$ in practice.
Additionally, the time complexity of each iteration of \kk is $\cO(n^2)$.
This means that the iterations of \kk are difficult to scale even for medium-sized datasets.
Therefore, although \kk provides higher quality clusterings than the classical \kmeans in certain cases, it scales poorly with respect to the size of the data set, limiting its practical applicability on CPU architectures.

\section{Matrix-Centric \kk}\label{sec:mtx-formulation}

This section describes our formulation of \kk in terms of sparse linear algebra primitives. 
First, we show that pairwise Euclidean distances between points and centroids can be computed using SpMM and matrix addition. 
Then, we describe strategies for efficiently computing the terms in this expression using sparse matrix computation.

\subsection{Computing Pairwise Distances}

Pairwise distances between points and centroids can be computed with matrix multiplication and matrix addition.
While prior work ~\cite{kernelmethods, baydoun2018cpu} has shown that this is possible, our work is the first to introduce the use of SpMM and SpMV in this computation.
In this section, the word `distance' refers to squared Euclidean distance; $\mathbf{p}_i$ and $\mathbf{c}_j$ are row vectors.

Let $\phi$ denote a non-linear function that maps a point $\mathbf{p}_i$ to a high dimensional feature space of dimensionality $\hat{d}$, and let $\kappa$ denote the corresponding kernel function.
Furthermore, let \K$\in \R^{n \times n}$ denote a matrix defined as follows:

\begin{equation}
    \mathbf{K} = \begin{bmatrix} \phi(\mathbf{p}_1)\phi(\mathbf{p}_1^T) & \hdots & \phi(\mathbf{p}_1)\phi(\mathbf{p}_n^T) \\
    \vdots & \ddots & \vdots \\
    \phi(\mathbf{p}_n)\phi(\mathbf{p}_1^T) & \hdots & \phi(\mathbf{p}_n)\phi(\mathbf{p}_n^T)
    \end{bmatrix}
\end{equation}

\noindent
The entries of \K can be calculated with the kernel trick since they are the inner products of the points projected into the high-dimensional feature space.
This means that \K$_{ij} = \kappa(\mathbf{p}_i, \mathbf{p}_j)$, i.e. \K can be computed using the kernel function. 

The distance between a single point $\mathbf{p}_i$ and a single centroid $\mathbf{c}_j$ in feature space is given by:

\begin{multline*}
\|\phi(\mathbf{p}_i) - \mathbf{c}_j\|^2 = \sum_{l=0}^{\hat{d}}{(\phi(\mathbf{p}_i)^{(l)} - \mathbf{c}_j^{(l)})^2} = \\
\sum_{l=0}^{\hat{d}}(-2\phi(\mathbf{p}_i)^{(l)}\mathbf{c}_j^{(l)} + [\phi(\mathbf{p}_i)^{(l)}]^2 + [\mathbf{c}_j^{(l)}]^2)
\end{multline*}.

\noindent
Distributing the summation term, we obtain:
\begin{multline}\label{eq:dist_1}
\sum_{l=0}^{\hat{d}}(-2\phi(\mathbf{p}_i)^{(l)}\mathbf{c}_j^{(l)} + [\phi(\mathbf{p}_i)^{(l)}]^2 + [\mathbf{c}_j^{(l)}]^2) = \\
\sum_{l=0}^{\hat{d}}-2\phi(\mathbf{p}_i)^{(l)}\mathbf{c}_j^{(l)} + \sum_{l=0}^{\hat{d}}[\phi(\mathbf{p}_i)^{(l)}]^2 +
\sum_{l=0}^{\hat{d}}[\mathbf{c}_j^{(l)}]^2 = \\
\sum_{l=0}^{\hat{d}}-2\phi(\mathbf{p}_i)^{(l)}\mathbf{c}_j^{(l)} + \|\phi(\mathbf{p}_i)\|^2 + \|\mathbf{c}_j\|^2.
\end{multline}

\noindent
Define two matrices, \Pmat$ \in \R^{n\times \hat{d}}$ and \C$ \in \R^{k\times \hat{d}}$ as:

\begin{multline}
\setlength{\arraycolsep}{3.5pt}
\begin{array}{ll}
{\mathbf{P}} = \begin{bmatrix}
   \phi(\mathbf{p}_1)^{(1)} & \hdots & \phi(\mathbf{p}_1)^{(\hat{d})} \\ 
   \vdots & \ddots & \vdots \\ 
   \phi(\mathbf{p}_n)^{(1)} & \hdots & \phi(\mathbf{p}_n)^{(\hat{d})}
\end{bmatrix}, &
{\mathbf{C}} = \begin{bmatrix}
   \mathbf{c}_1^{(1)} & \hdots & \mathbf{c}_1^{(\hat{d})} \\ 
   \vdots & \ddots & \vdots \\ 
   \mathbf{c}_k^{(1)} & \hdots & \mathbf{c}_k^{(\hat{d})}
\end{bmatrix}
\end{array}  
\end{multline}.

\noindent
In addition, define $\mathbf{\tilde{P}} \in \R^{n \times k}$ and $\mathbf{\tilde{C}} \in \R^{n \times k}$ as:

\begin{equation}
\setlength{\arraycolsep}{3.5pt}
\begin{array}{ll}
\mathbf{\tilde{P}} = \begin{bmatrix}
    \|\phi(\mathbf{p}_1)\|^2 & \|\phi(\mathbf{p}_1)\|^2 & \hdots & \|\phi(\mathbf{p}_1)\|^2 \\
    \vdots & \vdots & \ddots & \vdots \\ 
    \|\phi(\mathbf{p}_n)\|^2 & \|\phi(\mathbf{p}_n)\|^2 & \hdots & \|\phi(\mathbf{p}_n)\|^2
\end{bmatrix},\\
\mathbf{\tilde{C}} = \begin{bmatrix}
    \|\mathbf{c}_1\|^2 & \|\mathbf{c}_2\|^2 & \hdots & \|\mathbf{c}_k\|^2 \\ 
    \vdots & \vdots & \ddots & \vdots \\
    \|\mathbf{c}_1\|^2 & \|\mathbf{c}_2\|^2 & \hdots & \|\mathbf{c}_k\|^2
\end{bmatrix}.
\end{array}  
\end{equation}

\noindent
Computing the matrix-matrix product $\mathbf{-2PC^T}$ produces a matrix that stores $\sum_{l=0}^{\hat{d}}-2\phi(\mathbf{p}_i)^{(l)}\mathbf{c}_j^{(l)}$ for each $\mathbf{p}_i, \mathbf{c}_j$.
This accounts for the first term in Equation \ref{eq:dist_1}.
Adding $\mathbf{\tilde{P}}, \mathbf{\tilde{C}}$ to $\mathbf{-2PC^T}$ accounts for the $\|\phi(\mathbf{p}_i)\|^2, \|\mathbf{c}_j\|^2$ terms in Equation \ref{eq:dist_1}. 
Therefore, computing Equation 1 for each pair of $\mathbf{p}_i, \mathbf{c}_j$ is equivalent to the following:

\begin{equation} \label{eq:mtx_dist}
\mathbf{D = -2PC^T + \tilde{P} + \tilde{C}}.
\end{equation}

This equation is capable of computing pairwise distances between points and centroids, but it requires precisely computing the coordinates of the points in the high-dimensional feature space to initialize \Pmat and \C, which is difficult and computationally expensive \cite{kernelmethods}.
Furthermore, if $\hat{d}$ is large, the matrix-matrix product $\mathbf{PC^T}$ is also expensive.
Thus, we need to modify Equation \ref{eq:mtx_dist} so that it does not require the precise locations of the points in feature space.
Centroids are given according to the following expression:

\begin{equation}
\mathbf{c}_j = \frac{1}{|L_j|}\sum_{\mathbf{p}_i \in L_j}\phi(\mathbf{p}_i)
\end{equation}

This expression sums the points in the same cluster and divides the sum by the cardinality of the cluster.
One can also think of this operation as summing the rows of \Pmat that correspond to points in the same cluster and then dividing the summed rows by the appropriate cluster cardinality.

Then, define a matrix \V$ \in \R^{k \times n}$ as follows:

\begin{equation}\label{eq:v_init}
\mathbf{V}_{ji} = \begin{cases}
\frac{1}{|L_j|} & \mathbf{p}_i \in L_j \\
0 & \text{otherwise}
\end{cases}
\end{equation}

\V is a selection matrix that indicates which points are in which clusters.
It has one column for each point, and one row for each cluster. 
An entry is nonzero if the point corresponding to the entry's column index is in the cluster corresponding to the entry's row index.
In addition, note that \V is a sparse matrix with exactly $n$ nonzeros.

Using \V, new centroid locations are given by: 

\begin{equation}\label{eq:centroids}
\mathbf{C = VP}
\end{equation}
This matrix multiplication with \V can be thought of as selecting rows of \Pmat that correspond to points in the same cluster, dividing each of them by the cardinality of their cluster, and then adding them.
Using Equation~\ref{eq:centroids}, we can rewrite Equation~\ref{eq:mtx_dist} as:

\begin{equation}\label{eq:dist_sp_tmp}
\mathbf{D = -2PP^TV^T + \tilde{P} + \tilde{C} }
\end{equation}

\noindent
$\mathbf{PP^T}$ stores the pairwise inner products of the points projected into the high-dimensional feature space, or in other words, $\mathbf{(PP^T)}_{ij} = \phi(\mathbf{p}_i)\phi(\mathbf{p}_j^T)$.
This means that $\mathbf{PP^T = K}$, and we can rewrite Equation \ref{eq:dist_sp_tmp} as follows:

\begin{equation}\label{eq:dist_sp_1}
\mathbf{D = -2KV^T + \tilde{P} + \tilde{C} }
\end{equation}

$\mathbf{D}_{ij}$ gives the distance between the $i$th point and the $j$th centroid. 
We refer to $\mathbf{D}$ as the distances matrix in the rest of the paper.
This expression is able to calculate the distances between points and centroids using only matrix multiplication and matrix addition. Since this equation contains neither \Pmat nor \C, the precise coordinates of the points in the feature space do not need to be computed.
Section \ref{sec:norms} explains how to compute $\mathbf{\tilde{C}}$ without forming \C. 
In practice, since \V is a sparse matrix and \K is a dense matrix, the $-2\mathbf{K}\mathbf{V^T}$ term can be efficiently computed with SpMM.
This SpMM requires $\cO(n^2)$ work, as it involves the multiplication of a sparse matrix with $n$ nonzeros with a dense matrix that has $n$ columns.

\subsection{Computing the Kernel Matrix}

For common kernels, it is possible to compute \K with the help of matrix operations, which leads to an efficient implementation.
Here, we describe strategies to do this for two commonly used kernels: the polynomial kernel and the Gaussian kernel.
Let $\hat{\mathbf{P}} \in \R^{n \times d}$ be defined as 
\[
{\hat{\mathbf{P}}} = \begin{bmatrix}
   \mathbf{p}_1^{(1)} & \hdots & \mathbf{p}_1^{(d)} \\ 
   \vdots & \ddots & \vdots \\ 
   \mathbf{p}_n^{(1)} & \hdots & \mathbf{p}_n^{(d)}
\end{bmatrix}
\]
$\hat{\mathbf{P}}$ is a matrix that stores the points in their original input space.
This is in contrast to \Pmat, which stores the points projected into feature space. 

\noindent
The polynomial kernel is defined as:
\begin{equation*}
    \kappa(\mathbf{x}, \mathbf{y}) = (\gamma(\mathbf{x^Ty}) + c)^r
\end{equation*}
and the Gaussian kernel is defined as:
\begin{equation*}
    \kappa(\mathbf{x, y}) = \exp\left(-\frac{\gamma\|\mathbf{x - y}\|^2}{\sigma^2}\right)
\end{equation*}

For the polynomial kernel, computing \K is straightforward.
First, computing $\mathbf{B = \hat{P}\hat{P}^T}$ gives the pairwise inner products of the points.
From here, a sequence of element-wise operations on \B gives \K.
In particular,

\begin{equation}
\mathbf{K} = \text{pow}((\gamma\mathbf{B}) + c, r)
\end{equation}

\noindent
where $+$ is elementwise addition and $\text{pow}(\mathbf{X}, r)$ raises each element of the matrix $\mathbf{X}$ to the $r$th power. 

For the Gaussian kernel, computing \K is more complicated.
If we apply Equation \ref{eq:dist_1} to the definition of the Gaussian kernel, we get the following:

\begin{equation*}
    \kappa(\mathbf{x, y}) = \exp\left(\frac{-\gamma( -2\mathbf{x^Ty + x^Tx + y^Ty})}{\sigma^2}\right)
\end{equation*}
If we let $\mathbf{B=\hat{P}\hat{P}^T}$, then \B$_{ij} = \mathbf{p}_i\mathbf{p}_j^T$.
It follows that:

\begin{equation}\label{eq:gauss}
    \kappa(\mathbf{p}_i, \mathbf{p}_j) = \exp\left(\frac{-\gamma(-2\mathbf{B}_{ij} + \mathbf{B}_{ii} + \mathbf{B}_{jj})}{\sigma^2}\right)
\end{equation}
\K can therefore be calculated for the Gaussian kernel by first calculating $\mathbf{B=\hat{P}\hat{P}^T}$ and then using Equation~\ref{eq:gauss}.

Computing \B involves multiplying two dense matrices to produce a symmetric dense output matrix.
To do this, either the general matrix-matrix multiply (GEMM) BLAS routine or the symmetric rank-k update (SYRK) BLAS routine can be used.
We consider the performance-related implications of both approaches in Section \ref{sec:impl}.
Finally, note that \K does not change between iterations, so it is only necessary to compute \K once in \kk.

\subsection{Computing Row-Wise Norms}\label{sec:norms}

The row-wise norms of \Pmat and \C are needed to initialize the $\mathbf{\tilde{P}}$ and $\mathbf{\tilde{C}}$ matrices used in Equation \ref{eq:dist_sp_1}.
It is possible to efficiently compute these terms using sparse matrix operations without requiring substantial additional computation.
Let us first consider the calculation of the row-wise norms of \Pmat.
Recall from the previous section that $\mathbf{K=PP^T}$.
This implies that \K$_{ii} = \phi(\mathbf{p}_i)\phi(\mathbf{p}_i)^T$.
$\phi(\mathbf{p}_i)\phi(\mathbf{p}_i)^T = \|\phi(\mathbf{p}_i)\|^2$, so computing the row-wise norms of \Pmat can be done by extracting the diagonal of \K.
Since the calculation of \K is already required to calculate the $\mathbf{-2KV^T}$ term in Equation \ref{eq:dist_sp_1}, computing the row-wise norms of \Pmat does not require any extra computation.

Let us now consider the calculation of the row-wise norms of \C.
Recall that explicitly forming \C is not possible, since doing so would require explicitly projecting the points into feature space. 
Therefore, we need an expression for computing the norms of each row of \C that does not depend on \C itself.
Observe that, similarly to \Pmat, $(\mathbf{CC^T})_{ii} = \|\mathbf{c}_i\|^2$.
From Equation \ref{eq:centroids} we have that $\mathbf{C = VP}$.
It follows that $\mathbf{CC^T = VPP^TV^T = VKV^T}$, which finally results in:

\begin{equation}\label{eq:c_norm}
\mathbf{(VKV^T)}_{ii} = \|\mathbf{c}_i\|^2
\end{equation}

\noindent
The norms of \C can therefore be computed  without forming \C by calculating $\mathbf{VKV^T}$ and extracting the diagonal.

This approach produces a correct output, but does a lot of unnecessary work, as only the diagonal of $\mathbf{VKV^T}$ is needed.
To address this shortcoming, we design a new approach based on SpMV that computes only the diagonal of $\mathbf{VKV^T}$.

This approach takes advantage of a unique property of \V, namely that it has \textit{exactly one non-zero per column}.
To see why this is the case, recall that \V$_{ij} \neq 0$ if $\mathbf{p}_j \in L_i$.
Each $\mathbf{p}_j$ can only be in one cluster, which means that for each $j$, \V$_{ij} \neq 0$ can only apply to a single row index $i$.
This also implies that each non-zero in \V has a unique column index.
This means that the set of inner products that compute $\|\mathbf{c}_1\|^2 \hdots \|\mathbf{c}_k\|^2$ involve multiplication by disjoint nonzeros of $\mathbf{KV^T}$.
In other words, the nonzeros in the $i$th column of $\mathbf{KV^T}$ that are multiplied by the $i$th row of \V to form $\|\mathbf{c}_i\|^2$ are not involved in any of the other inner products that compute other $\|\mathbf{c}_j\|^2, j\neq i$.

\begin{figure}[t]
    \centering
    \includegraphics[width=\columnwidth]{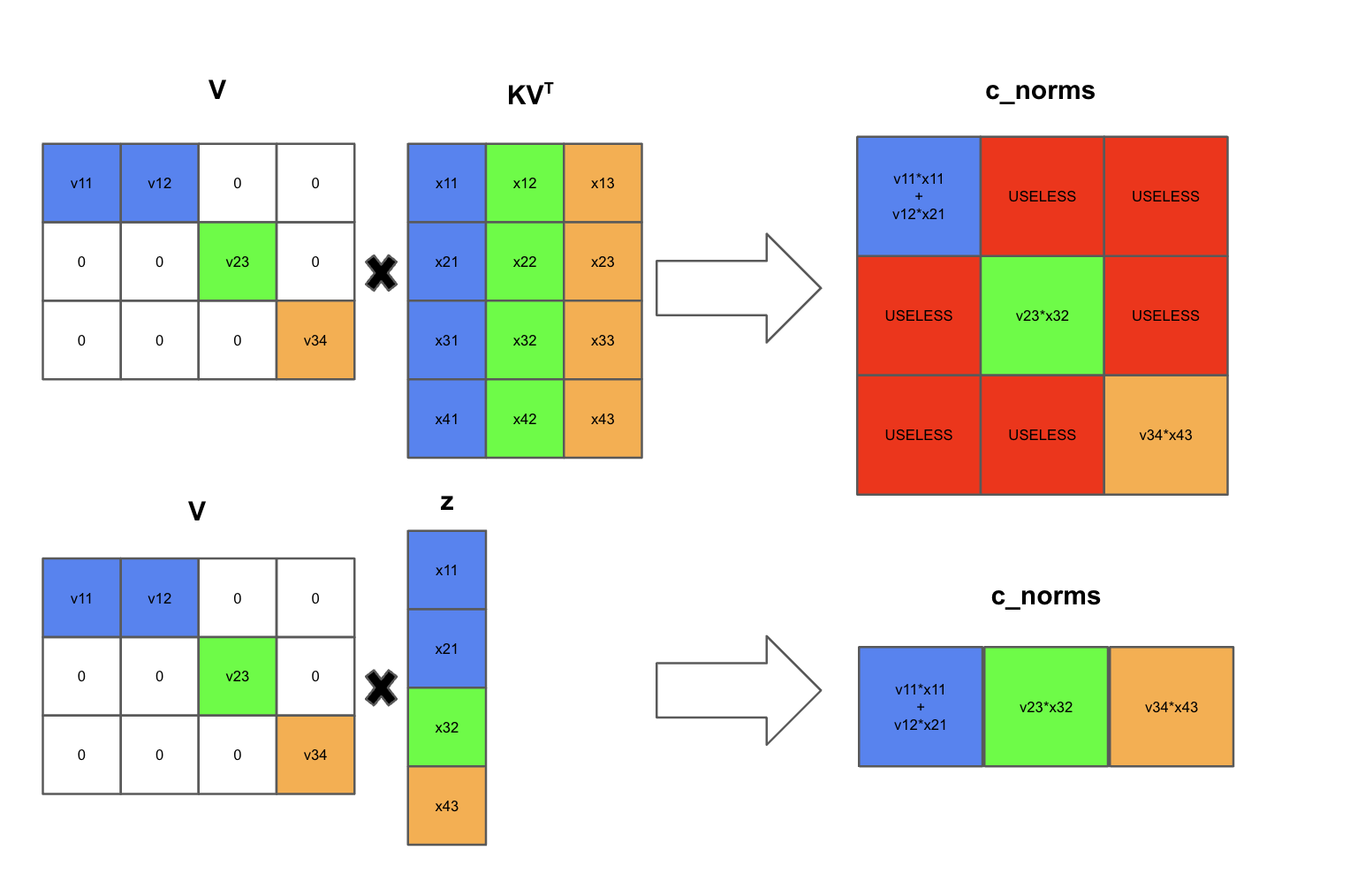}
    \caption{Computing the diagonal of $\mathbf{VKV^T}$ using SpMV.}
    \Description{}
    \label{fig:cnorms}
\end{figure}

Figure \ref{fig:cnorms} shows a simple hypothetical example of multiplication between \V and $\mathbf{KV^T}$.
The rows of \V and the columns of $\mathbf{KV^T}$ are color-coded to indicate which inner products between rows and columns are \textit{strictly} necessary to produce entries of $\mathbf{\tilde{C}}$.
If the entries of $\mathbf{KV^T}$ that are involved in these strictly necessary inner products are extracted into a dense vector \z, then the matrix-vector product $\mathbf{Vz}$ will only produce the entries needed to compute $\mathbf{\tilde{C}}$.

Formally, we can define the dense vector \z$ \in \R^n$:

\begin{equation}
    \mathbf{z} = \begin{bmatrix} 
    \mathbf{KV^T}_{1, \text{cluster}(1)} \\
    \mathbf{KV^T}_{2, \text{cluster}(2)} \\
    \vdots \\
    \mathbf{KV^T}_{n, \text{cluster}(n)}
    \end{bmatrix}
\end{equation}

Here, cluster($i$) is a function that returns the index of the cluster to which $\mathbf{p}_i$ is assigned.
Using these definitions, the norms of the centroids are given by:

\begin{equation}\label{eq:cnorms_gemv}
    \mathbf{Vz} = \begin{bmatrix}
    \|\mathbf{c_1}\|^2 \\
    \vdots \\
    \|\mathbf{c_k}\|^2 
    \end{bmatrix}
\end{equation}



\V is a sparse matrix with exactly $n$ nonzeros, so this matrix-vector product can be computed with SpMV while only requiring $\cO(n)$ work.
The previous approach of explicitly calculating $\mathbf{VKV^T}$ and extracting the diagonal requires $\cO(nk)$ work.
Note that $\mathbf{KV^T}$ is already computed in the first term of Equation~$\ref{eq:dist_sp_1}$, meaning $z$ can be initialized without any additional computation.
The only additional computation required in our algorithm for computing the norms of \C is the SpMV operation used to calculate $\mathbf{Vz}$.

Using Equation ~\ref{eq:cnorms_gemv}, the row-wise norms of \C can be computed \textit{without} explicitly forming the \C matrix.
This eliminates the need to compute new centroids at each iteration.
The only thing that changes with each iteration is the sparsity structure of \V.

\subsection{\popcorn Algorithm}

Here, we present the complete algorithm for our matrix-centric formulation of \kk, namely \popcorn.
Algorithm \ref{alg:kkmeans-mtx} shows pseudocode depicting the complete \popcorn algorithm.
Line 1 of the algorithm computes the kernel matrix \K.
This is done by first computing $\mathbf{B=\hat{P}\hat{P}^T}$ and then applying the chosen kernel to update each entry of \B.
Line 2 computes the row-wise norms of \Pmat, which are needed to initialize $\mathbf{\tilde{P}}$, by extracting the diagonal of \K.
As the points do not change between the iterations, \K is calculated once and $\mathbf{\tilde{P}}$ is initialized once.
Line 3 assigns each point in the dataset a random integer between $1$ and $k$, i.e. each point is assigned a random cluster label.
These are used in line 4 to initialize the first \V matrix according to Equation \ref{eq:v_init}.

The main loop of \kk begins on line 6.
Each iteration involves three steps: computing $\mathbf{-2KV^T}$ using SpMM, computing the norms of \C using SpMV, and updating cluster assignments.
Line 7 computes the scaled matrix-matrix product $\mathbf{-2KV^T}$ with SpMM.
Then, to calculate the row-wise norms of \C needed to initialize $\mathbf{\tilde{C}}$, line 8 initializes the dense vector \z with $\mathbf{-2KV^T}$ and cluster assignment of each point.
From here, line 9 computes the matrix-vector product $\mathbf{-0.5Vz}$ using SpMV, and the resulting vector is used to initialize $\mathbf{\tilde{C}}$.
The $-0.5$ scalar term is needed to cancel out the $-2$ used to scale $\mathbf{KV^T}$.
Finally, the actual distances are calculated in line 10 by adding the matrices $\mathbf{-2KV^T}$, $\mathbf{\tilde{P}}$ and $\mathbf{\tilde{C}}$.

The cluster assignments are updated via the loop beginning on line 12 by calculating the index of the smallest element in each row of the distance matrix \D.
This determines the centroid to which each point is closest, i.e. to which cluster each point should be reassigned.
Once this is done, \V is updated in line 14 to reflect the new assignments, and the iteration is complete.
This is repeated until the convergence, or until a maximum number of iterations have been executed.

\begin{algorithm}
\caption{\popcorn Algorithm for \kk}
\label{alg:kkmeans-mtx}
\begin{algorithmic}[1]
\State \K $= \texttt{kernel}(\mathbf{\hat{P}\hat{P}^T})$
\State Initialize $\mathbf{\tilde{P}}$ using $\diag(\mathbf{K})$
\State Assign each $p_1 \hdots p_n$ a random cluster
\State Initialize \V using cluster assignments
\State Initialize iteration counter $iter = 0$
\While{$iter < maxiter$ and centroids $c_j$ are changing}
    \State $\mathbf{E} = -2*\texttt{SpMM}(\mathbf{KV^T})$
    \State $\mathbf{z^T} = -0.5\begin{bmatrix}
    \mathbf{E}_{1, \text{cluster(1)}} & 
    \mathbf{E}_{2, \text{cluster(2)}} &
    \hdots &
    \mathbf{E}_{n, \text{cluster(n)}}
    \end{bmatrix}$
    \State Initialize $\mathbf{\tilde{C}}$ using $\texttt{SpMV}(\mathbf{Vz})$
    \State $\mathbf{D} = \mathbf{E} + \mathbf{\tilde{P}} + \mathbf{\tilde{C}}$
    \For{$i=1:n$}
        \State cluster(i) = $\texttt{argmin}_j(\mathbf{D}_{i,j})$ 
    \EndFor
    \State Set \V using cluster assignments 
    \State $iter \gets iter + 1$
\EndWhile
\end{algorithmic}
\end{algorithm}

\section{\popcorn Implementation}\label{sec:impl}

In this section, we provide details on our implementation of \popcorn.
Our implementation is designed for modern NVIDIA GPUs and is mainly based on the cuBLAS and cuSPARSE libraries, making light use of thrust \cite{Bell2012ThrustAP}.

\subsection{Data Preparation}\label{sec:data_pre}

There are two preparation stages.
First, we read the input data from disk into host memory and copy it into a device buffer.
The device buffer is a dense matrix $\mathbf{\hat{P}}$ that stores the data points in row-major order, i.e., $\mathbf{\hat{P}}_{i,:} = \mathbf{p}_i$.
Then, we randomly assign a cluster label between $1$ and $k$ to each point and use this random assignment to initialize the first \V matrix.
\V is stored using the Compressed Sparse Row (CSR) format provided by cuSPARSE.
CSR stores a sparse matrix using 3 arrays, a \texttt{values} array that stores the nonzeros, a \texttt{colinds} array that stores the column indices of the nonzeros, and a \texttt{rowptrs} array that stores the start and end indices of each row in the other two arrays.
The cluster assignments are first created and stored in an array on the device.
Then, the cardinality of each cluster is calculated via a reduction of the cluster assignment array.
Finally, \V is initialized with a hand-written CUDA kernel that uses the cluster assignment and cardinality arrays to set the CSR data structures.

\subsection{Computing the Kernel Matrix}

Before \kk can be executed, the kernel matrix \K must be computed.
This is done by first computing $\mathbf{B = \hat{P}\hat{P}^T}$. 
One of two cuBLAS routines can be used to calculate $\mathbf{B}$, either general matrix-matrix multiplication (GEMM) or symmetric rank-k update (SYRK).
Both routines provide correct output, but when deciding between GEMM and SYRK, there are some performance-related trade-offs to consider.
$\mathbf{B}$ is symmetric, so it is only strictly necessary to compute either its upper or lower triangular region. 
SYRK explicitly calculates only one of the two triangular areas, while GEMM calculates the entirety of \B.
Therefore, GEMM does unnecessary work.
More precisely, using GEMM to compute \B requires $\cO(n^2d)$ FLOPS, while using SYRK requires only $\cO(\frac{n^2d}{2})$ FLOPS.
Despite using fewer FLOPS, if SYRK is used to compute \B, it is necessary to copy the triangular area of \B which is calculated explicitly into the triangular area of \B which is not calculated explicitly.
This is because the cuSPARSE routines used in \popcorn require the entirety of \B to be stored explicitly, even if it is symmetric.

Depending on the values of $n$ and $d$, it may be more efficient to choose one routine over the other.
In practice, we find that for input data with large values of $n$ and small values of $d$, using GEMM to compute \B is more efficient while using SYRK to compute \B is more efficient when $d$ is closer to or larger than $n$.
To ensure that the most efficient routine is used, \popcorn computes the ratio $r$ between $n$ and $d$.
If $r$ exceeds a certain threshold $t$, GEMM is used, otherwise SYRK is used.
The appropriate value of $t$ is architecture-dependent, so we leave it as a tunable parameter.
In Section~\ref{sec:results}, we investigate the appropriate value of $t$ for a specific architecture.

Once \B has been computed using either the GEMM-based algorithm or the SYRK-based algorithm, the kernel function is applied to each entry of \B using \texttt{thrust::transform}, which produces the final kernel matrix \K.
The diagonal of \K is then extracted and stored in a dense vector of length $n$ on the device.
This vector implicitly represents the entire $\tilde{\mathbf{P}}$ matrix, since each column of $\tilde{\mathbf{P}}$ is identical.
This part of the implementation corresponds to lines 1-4 of Algorithm \ref{alg:kkmeans-mtx}.
Once \K and $\tilde{\mathbf{P}}$ are initialized, the main iterations of \kk can begin.

\subsection{Pairwise Distance Computation}

The most costly part of each iteration involves computing the pairwise distances between points and centroids using Equation~\ref{eq:dist_sp_tmp}.
First, $-2\mathbf{KV}^T$ is computed using the cuSPARSE SpMM routine.
Then, \z is initialized using a hand-written CUDA kernel that uses an array containing cluster assignments to set each element of \z to the appropriate entry of $-2\mathbf{KV}^T$.
The matrix-vector product $-0.5\mathbf{Vz}$ is then calculated with the cuSPARSE SpMV routine.
The output is a dense vector used to implicitly represent the entire $\tilde{\mathbf{C}}$ matrix.
As with $\tilde{\mathbf{P}}$, the rows of $\tilde{\mathbf{C}}$ are identical so it is not necessary to store the entire matrix.

From here, the distances matrix \D can be computed by summing $\mathbf{-2KV^T}, \tilde{\mathbf{P}}$, and $\tilde{\mathbf{C}}$.
Note that, since $\tilde{\mathbf{P}}$ and $\tilde{\mathbf{C}}$ are implicitly represented as vectors, a custom kernel is necessary to add them to $-2\mathbf{KV}^T$.
This custom kernel uses one thread to update each entry of $-2\mathbf{KV}^T$ with the appropriate entries of the vectors used to represent $\tilde{\mathbf{P}}$ and $\tilde{\mathbf{C}}$.
The corresponding entry of the $\tilde{\mathbf{P}}$ vector that each thread should access can be computed by dividing the global ID of the thread by $k$, and the same can be done for $\tilde{\mathbf{C}}$ by taking the global ID of the thread modulo $k$.
Once the matrices have been added, the distances matrix \D is computed.
This part of the implementation is described by lines 7-10 of Algorithm \ref{alg:kkmeans-mtx}.

Cluster assignment is relatively inexpensive for \kk, so matrix computations are not used. 
It is done by computing the index of the smallest element in each row of \D using the \texttt{coalescedReduction} function from NVIDIA's RAPIDS \cite{rapidsai} library. 
Once assignments are ready, \V is updated using the procedure in Section \ref{sec:data_pre}.
This corresponds to lines 11-14 of Algorithm \ref{alg:kkmeans-mtx}.

\subsection{Arithmetic Intensity}

Here we give the arithmetic intensity of \popcorn's algorithm for \kk.
We only consider the arithmetic intensity of the operations used to compute \K and \D, as other components of the algorithm do not involve significant amounts of arithmetic.
We assume all matrices and vectors are stored using single-precision floating point numbers, and we further assume \V is stored using 32 bit indices.

Computing the kernel matrix \K requires multiplying a dense matrix of size $n \times d$ by its transpose to produce \B, and then applying the kernel function to produce \K.
Let $F_K, B_K$ respectively denote the number of FLOPS and memory operations needed to use the kernel function to to produce \K.
The arithmetic intensity of computing \K is then 
\begin{equation}
    \frac{F_K +2n^2d}{4(B_K +2nd + n^2)}
\end{equation}
%
Computing \D requires one SpMM, one SpMV, and elementwise addition of three dense matrices.
Recall that two of these dense matrices, $\mathbf{\tilde{P}}$ and $\mathbf{\tilde{C}}$, are stored as dense vectors of lengths $n$ and $k$, respectively. Therefore, the total arithmetic intensity of computing \D each iteration is
\begin{equation}
    \frac{2n^2 + 2n + 3nk}{4(n^2 + 6n + 4k + 3nk)} 
\end{equation}

\subsection{Benefits of Using Sparse Matrices}

The key feature of \popcorn is its use of sparse matrix computations to process expensive components of \kk.
This has several advantages.

\paragraph{Ease of Programmability}

GPU-based applications are usually developed using a programming framework such as CUDA.
The manual kernel engineering, tuning, and performance optimization required to achieve good performance for most CUDA codes is significant and time- consuming \cite{cudaHard}.
This is because programming effectively with CUDA requires the programmer to reason about many intricate details of the low-level behavior of GPUs and explicitly manage shared memory, grid sizes, and block sizes.
In addition, kernels must be written to achieve good load balancing between thread blocks.

\popcorn did not require extensive manual kernel engineering effort. 
This is because our formulation based on SpMM and SpMV makes it possible to offload most of the computation to in \kk library routines, eliminating the need to write and optimize large CUDA kernels by hand.
\popcorn uses a small number of hand-written kernels, but these kernels are simple and involve straightforward grid configurations and are embarassingly parallel.

\paragraph{Guaranteed High Performance} 

Offloading computation to cuSPARSE and cuBLAS provides a reasonable guarantee of good performance.
 cuSPARSE and cuBLAS routines are highly optimized, and substantial effort has been made to ensure that these libraries use GPU resources effectively.
Thanks to the use of these libraries, \popcorn can utilize the high performance of cuSPARSE and cuBLAS.
Furthermore, this high performance is portable between different NVIDIA GPU architectures and future improvements to cuSPARSE and cuBLAS will automatically lead to performance improvements in \popcorn.

%


\section{Results}\label{sec:results}

In this section, we show that \popcorn achieves a significant speedup over a GPU implementation of \kk. 
First, we examine the performance of different strategies for calculating the kernel matrix \K with different values of $n$ and $d$.
Then, we show the speedup of \popcorn's algorithm for computing pairwise distances over a baseline CUDA implementation.
Finally, we show the speedup of the entire algorithm of \popcorn compared to a baseline CUDA implementation.

\subsection{Experimental Setup}

This section contains details of the datasets and hardware used to test the performance of \popcorn.

\subsubsection{Hardware and Software Environment}

The experiments were performed on an NVIDIA A100 GPU connected to a 64-core AMD EPYC 7763 CPU via a PCIe Generation 4 interconnect.
All codes were compiled with NVCC 12.2 and use the versions of cuSPARSE, cuBLAS, and thrust from the CUDA Toolkit 12.2.

\subsubsection{Datasets} 

To test \popcorn, we used six real-world datasets from the libSVM data repository~\cite{chang2011libsvm}.
These datasets represent tasks from the fields of computer vision, image processing, natural language processing, and vehicle classification.
Additionally, the MNIST and CIFAR-10 datasets have been shown to benefit from using \kk to find non-linearly separable clusters \cite{6737057, large}.
The values of $d$ (number of features) and $n$ (number of samples) for each dataset can be seen in Table~\ref{tab:datasets}.

\subsubsection{Other Info}

Unless otherwise stated, the numbers presented in all experiments are average values over 4 trials.
In addition, for experiments with clustering runtimes, all implementations were run for exactly 30 iterations to ensure that differences in runtimes were not due to differences in convergence.
For each dataset, we set $k = \{10, 50, 100\}$ to investigate how \popcorn behaves given different cluster granularities.
The choice of kernel function does not influence the runtime, since most common kernel functions involve applying a simple function to each entry in the kernel matrix.
In this work, we use the polynomial kernel with $\gamma=1, c=1, d=2$.
Determining a suitable kernel function for a given dataset is outside the scope of this paper, and is discussed in detail elsewhere~\cite{large, kernelmethods}.

\begin{table}[t!]
    \centering
    \caption{Information about datasets used in experiments.}
    \resizebox{\columnwidth}{!}{
    \begin{tabular}{|c|c|c|c|}
        \hline 
         \textbf{Dataset} & \textbf{Description} & \textbf{n} & \textbf{d}  \\
         \hline
         Acoustic & Vehicle sensor data & 78823 & 50  \\
         \hline 
         CIFAR-10 & 32x32 color images & 50000 & 3072 \\
         \hline
         Ledgar & Large corpus of legal documents & 70000 & 19996 \\
         \hline 
         Letter & Hand-written letters & 10500 & 26 \\
         \hline 
         MNIST & Hand-written digits dataset & 60000 & 780 \\
         \hline 
         SCOTUS & Text of US Supreme Court rulings & 6400 & 126405 \\
         \hline
    \end{tabular}
    } 
    \vspace{1em}
    \label{tab:datasets}
    \vspace{-1em}
\end{table}

\begin{figure}[t!]
    \centering
    \includegraphics[width=\columnwidth]{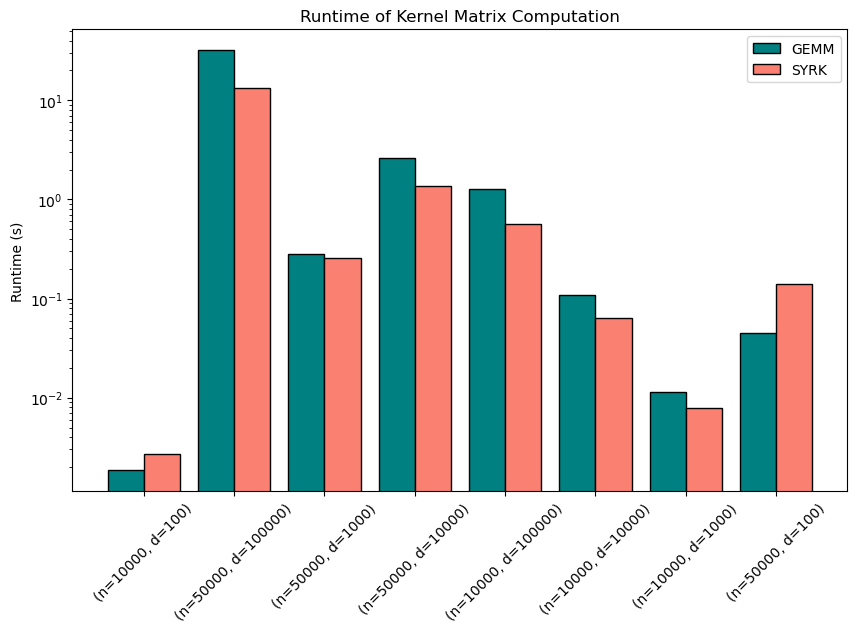}
    \caption{Comparison of the kernel matrix computation for synthetic data with SYRK and with GEMM.}
    \Description{}
    \label{fig:kernel-mtx-comp}
\end{figure}

\subsection{Kernel Matrix Computation}

As described in Section \ref{sec:impl}, \popcorn has two different algorithms that can be used to calculate the kernel matrix \K: the GEMM-based algorithm and the SYRK-based algorithm.
Here, we examine the performance of both strategies on synthetic datasets with different values of $n$ and $d$; we chose to use synthetic data to more thoroughly investigate the impact of varying $n$ and $d$.

Figure~\ref{fig:kernel-mtx-comp} shows the runtimes of GEMM and SYRK-based algorithms on synthetic data with different values of $n$ and $d$.
First, we set $n$ to $50000$ and $10000$ to simulate large and small dataset cardinalities, respectively, and for each value of $n$ we set $d$ to $100, 1000, 10000,$ and $100000$ to simulate different dimensionalities.

For datasets with large $n$ and small $d$, the GEMM-based algorithm is faster than the SYRK-based algorithm.
This is most significant for $(n=50000, d=100)$, where the GEMM-based algorithm is 3.2$\times$ faster than the SYRK-based algorithm.
For datasets where $n$ and $d$ are similar, the SYRK-based algorithm is up to 2.4$\times$ faster.
Overall, these experiments suggest that for our specific platform, it is best to use the GEMM-based algorithm when the ratio between $n$ and $d$ is greater than 100, while it is best to use the SYRK-based algorithm when the ratio is less than 100.
The reason for this is the overhead associated with copying the triangular half of \K which is explicitly computed into the triangular half of \K which is not explicitly computed.
If it were possible to perform SpMM and SpMV with a dense matrix stored in triangular packed form, this copy would not be necessary and the SYRK-based algorithm could be faster.



\subsection{CUDA Baseline Implementation}

To demonstrate the advantages of \popcorn's sparse matrix-based approach, we compare the time to solution of \popcorn's algorithm with a baseline CUDA implementation that does not utilize sparse matrix computations.
To the best of the authors' knowledge, there is no open-source GPU implementation of \kk with which we could have compared.
Baydoun et al~\cite{baydoun2018cpu} describe a GPU implementation of \kk, but it is not open-source, and although we contacted the authors, we were unable to access the code.
Given the differences between the time complexity of classical \kmeans and \kk, a comparison between \popcorn and a GPU implementation of classical \kmeans would not be meaningful.

Given the lack of a GPU-based open-source implementation of \kk, we compare \popcorn with an in-house CUDA implementation that does not use sparse matrix operations.
This baseline CUDA implementation uses the GEMM routine provided by cuBLAS to compute the kernel matrix and then uses three hand-written CUDA kernels to compute the distances between points and centroids.

The first kernel is responsible for reducing the entries of \K corresponding to points in the same cluster.
This kernel creates $n$ thread blocks and has each one iterate through a row of \K.
The entries of the row are reduced based on the cluster assignment of the point corresponding to the column index of the entry.
To reduce global memory accesses, the entries of each row are reduced into a buffer in shared memory of length $k$.
Once the rows have been reduced, the first $k$ threads in each thread block write the entries of the shared memory buffer into global memory in an embarrassingly parallel manner.
This kernel performs the same function as the SpMM in \popcorn, and it dominates the runtime of the baseline implementation.

The second kernel uses the reduced entries of \K computed by the first kernel to compute the row-wise norms of the centroids.
This kernel creates $n$ threads, each of which indexes the reduced buffer computed by the first kernel according to the cluster assignment of its corresponding point.
Entries of the reduced buffer corresponding to points in the same cluster are reduced and then written to the global memory in an embarrassingly parallel fashion.
This produces the row-wise norms of the centroids, performing the same function as the SpMV in \popcorn.

The last kernel calculates the complete distances between points and centroids using the buffers computed by the first two kernels.
This kernel creates $nk$ threads and has each one compute the distance between a single point and a single centroid by summing one entry from the two buffers computed by the first two kernels with an entry from the diagonal of \K.
No synchronization or inter-thread coordination is required as the operation is embarrassingly parallel.

\subsection{CUDA Baseline Comparison to CPU}\label{sec:cpu-comparison}

The goal of this section is to show that our baseline CUDA implementation is faster than a CPU implementation and therefore provides a reasonable baseline for comparison with \popcorn.
In the experiments, we measure the time required to compute the kernel matrix \K (line 1 in Algorithm \ref{alg:kkmeans-mtx}) and the runtime of the clustering itself (lines 6-16 of Algorithm \ref{alg:kkmeans-mtx}) and compare it to the implementation of \kk in the PRMLT package~\cite{chen2024prmlt}.
This implementation is a single-threaded CPU version of \kk.

\begin{figure}[t]
    \centering
    \includegraphics[width=\columnwidth]{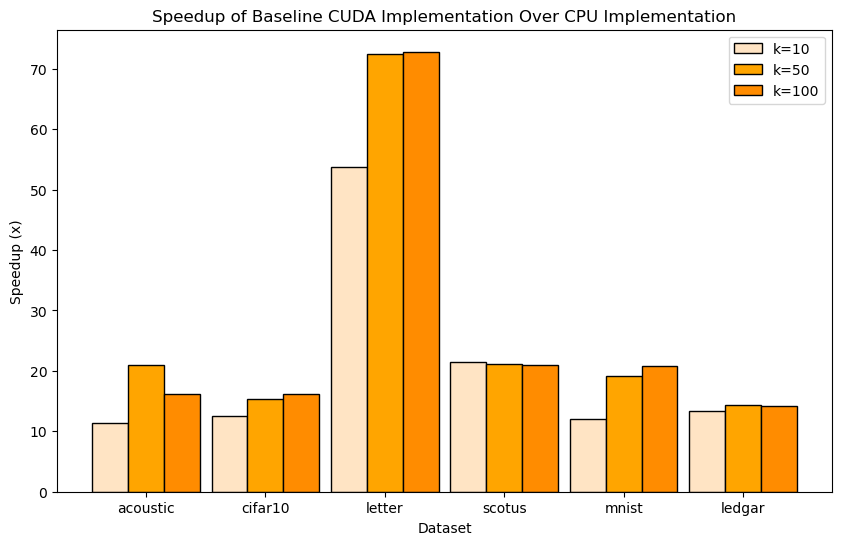}
    \caption{Baseline CUDA implementation speedup over CPU varying $k$.}
    \Description{}
    \label{fig:cuda}
\end{figure}

Figure~\ref{fig:cuda} illustrates the speedup of our baseline CUDA implementation compared to PRMLT \kk for each dataset.
Our baseline CUDA implementation consistently outperforms the CPU-only implementation.
The speedup is most noticeable for the letter dataset, which is a maximum of $72.8\times$ faster than the CPU-only implementation.
In general, the speedup ranges from $ 11\times$ to$21\times$, and tends to be higher for $k=50$ and $k=100$ than for $k=10$.
This is because, for smaller values of $k$, the load imbalance between thread blocks tends to be larger as the cluster sizes tend to be unbalanced.
Overall, our baseline CUDA implementation is at least an order of magnitude faster than the CPU-only implementation for all datasets and for all values of $k$.


\subsection{Pairwise Distances Computation}\label{sec:dist-perf}

Here, we compare \popcorn's SpMM and SpMV-based algorithm for computing pairwise distances to that of the baseline CUDA implementation described in \ref{sec:cpu-comparison}.
Importantly, this section does not consider the time required to compute the kernel matrix \K, since our goal in this section is to investigate the impact of using SpMM and SpMV on performance when computing pairwise distances, which is independent of computing \K.
Thus, we consider the speedup of \popcorn's distances algorithm over the baseline CUDA algorithm and the throughput of both algorithms.
In addition, we present a roofline model \cite{roofline} that shows how close both algorithms come to their respective theoretical peak throughput.

\begin{figure}[t]
     \centering
     \includegraphics[width=\columnwidth]{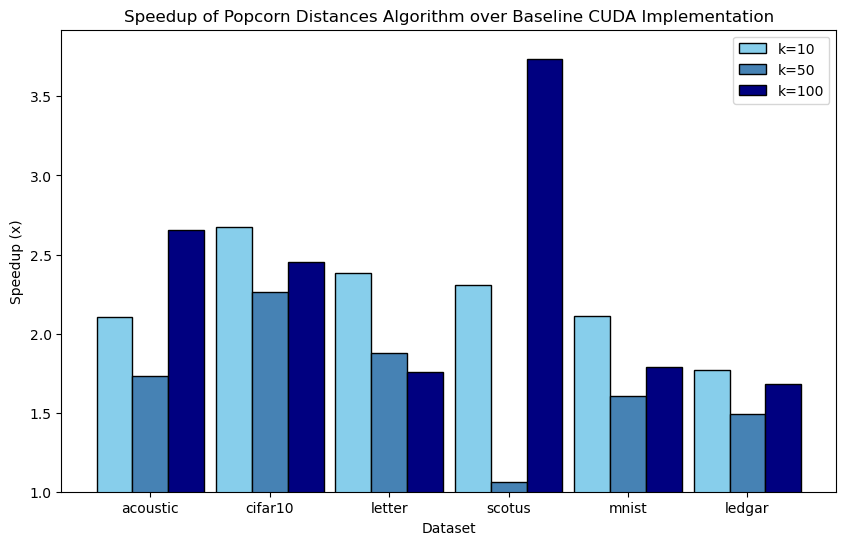}
     \caption{Speedup of \popcorn's pairwise distances algorithm over the baseline CUDA implementation varying $k$.}
    \Description{}
     \label{fig:distances-speedup}
 \end{figure}

Figure~\ref{fig:distances-speedup} shows the speedup of \popcorn's pairwise distances algorithm over that of the baseline CUDA implementation.
\popcorn is consistently between $1.5\times$ to $ 2.6\times$ faster than the baseline CUDA implementation, except for the SCOTUS dataset at $k=50$, where the speedup is $1.1\times$ due to the small number of points in this dataset, i.e., $n=6400$.

\begin{figure}[t]
    \centering
    \includegraphics[width=\columnwidth]{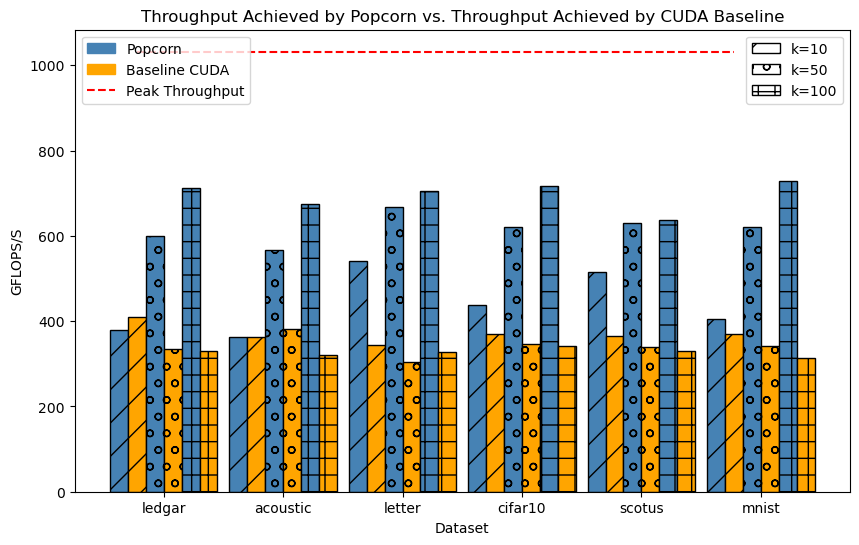}
    \caption{Comparison of throughput between the pairwise distances algorithm of \popcorn and the baseline CUDA implementation for varying $k$.}
    \Description{}
    \label{fig:popcorn-throughput}
\end{figure}

Figure \ref{fig:popcorn-throughput} illustrates the throughput in GFLOPS/s of the pairwise distances algorithm of \popcorn next to the throughput of the baseline implementation algorithm. 
The throughput values were obtained using Nsight Compute\cite{ncu}, and the peak throughput series in the figure was also obtained using Nsight Compute.
In \popcorn, we only measure the throughput of the SpMM operation, as this dominates the runtime, and similarly in the baseline implementation, we only measure the throughput of the first CUDA kernel.

\popcorn consistently achieves a higher throughput than our baseline implementation, ranging from 370 GFLOPS/s to 729 GFLOPS/s.
The baseline implementation achieves a throughput of between 304 GFLOPS/s and 409 GFLOPS/s.
In addition, \popcorn achieves progressively higher throughput with the scaling of $k$, while the baseline implementation achieves lower throughput with the scaling of $k$.
These results show that \popcorn can utilize GPU resources more effectively than hand-written CUDA kernels by using sparse matrix computation.

\begin{figure*}[t]
    \centering
    \begin{subfigure}[b]{0.3\textwidth}
        \centering
        \includegraphics[width=\textwidth]{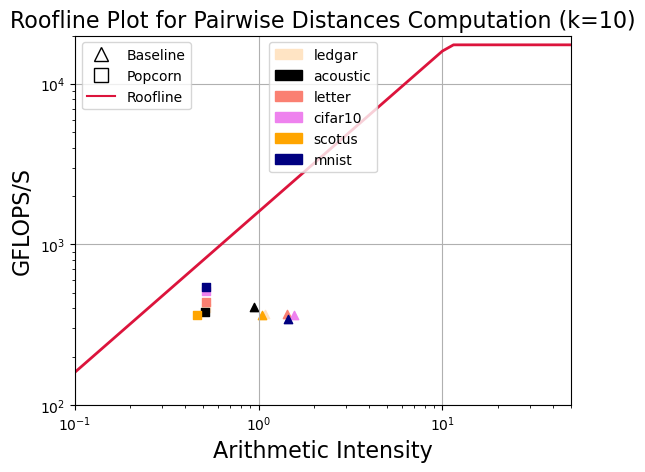}
    \end{subfigure}
    \begin{subfigure}[b]{0.3\textwidth}
        \centering
        \includegraphics[width=\textwidth]{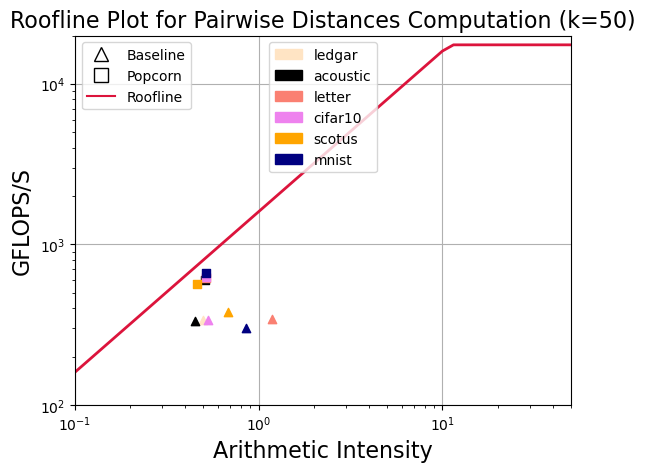}
    \end{subfigure}
    \begin{subfigure}[b]{0.3\textwidth}
        \centering
        \includegraphics[width=\textwidth]{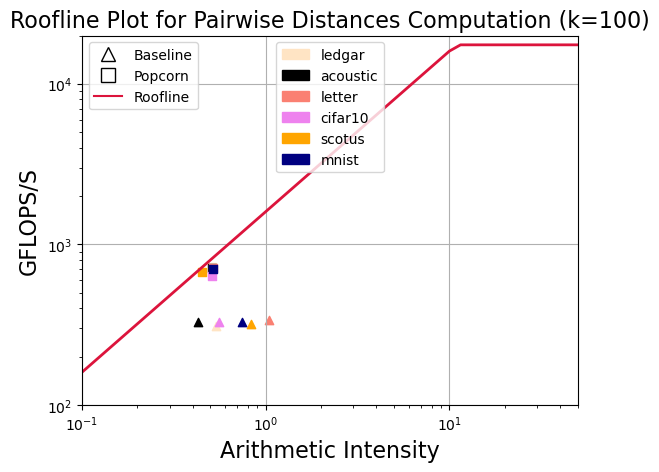}
    \end{subfigure}
    \caption{Roofline plots comparing pairwise distances algorithm of \popcorn and the baseline CUDA implementation.}
    \Description{}
    \label{fig:popcorn-roofline}
\end{figure*}

Figure \ref{fig:popcorn-roofline} illustrates the roofline model for each dataset and each value of $k$.
In these plots, the triangles represent the throughput achieved with the baseline implementation and the squares represent the throughput achieved with \popcorn.
The arithmetic intensity values were collected using Nsight Compute.
\popcorn is consistently closer to the roofline than the baseline implementation for all datasets, and for $k=\{50, 100\}$ \popcorn almost hits the roofline in all cases.
Notably, \popcorn has a lower arithmetic intensity for some data sets than the baseline implementation.
By profiling with Nsight Compute, we found out that this is because our baseline implementation reduces the entries of the kernel matrix in shared memory, while the SpMM routine of cuSPARSE and thus also \popcorn does not use shared memory.
As a result, the off-chip memory traffic for \popcorn is larger, hence the lower arithmetic intensity.

\subsection{\popcorn Comparison to Baseline}\label{sec:runtime_comparison}

In this section, we compare \popcorn with the baseline CUDA implementation.
In these experiments, we measure the time required to compute \K as well as the runtime of the clustering itself.
\popcorn uses the SYRK-based algorithm to compute the kernel matrix \K if $\frac{n}{d} < 100$, otherwise it uses the GEMM-based algorithm.

\begin{figure}[t]
    \centering
    \includegraphics[width=\columnwidth]{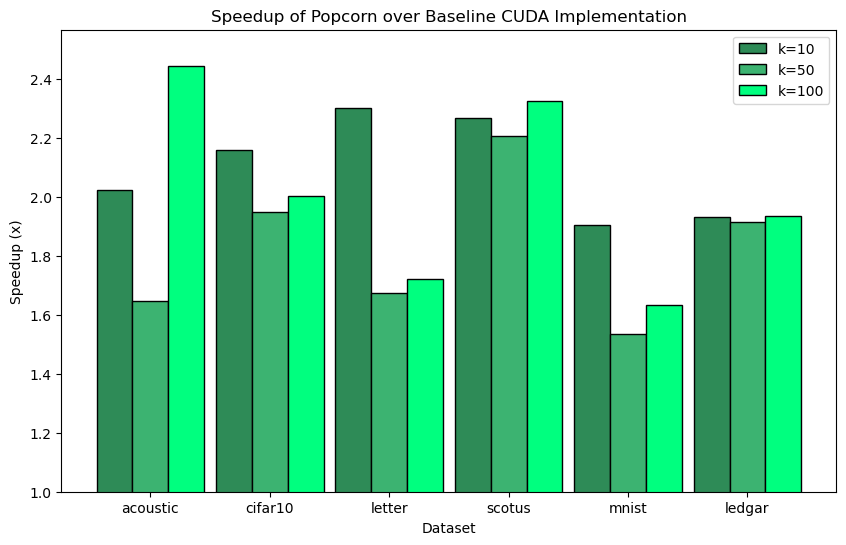}
    \caption{\popcorn speedup over baseline CUDA implementation varying $k$.}
    \Description{}
    \label{fig:popcorn-speedup}
\end{figure}

Figure~\ref{fig:popcorn-speedup} illustrates the speedup of \popcorn over the baseline CUDA implementation for each test dataset for different values of $k$.
\popcorn is consistently faster than the baseline CUDA implementation. 
For $k=10$, the speedup ranges from $ 1.9\ times$ to $ 2.3\ times$.
For $k=50$ and $k=100$, the speedup is similar and ranges from $1.6\times$ to $2.4\times$ for $k=50$, and from $1.6\times$ to $2.6\times$ for $k=100$.
Overall, \popcorn is consistently faster than our baseline CUDA implementation.
These speedups are due to a combination of \popcorn's strategy for selecting the optimal kernel for computing \K and its superior use of GPU resources thanks to SpMM and SpMV.


\subsection{Runtime Breakdown}

In this section, we present a runtime breakdown of \popcorn's algorithm for \kk.
We consider the time required to compute the kernel matrix \K, the time required to compute the pairwise distances using SpMM and SpMV, and the time required to update the cluster assignments.

\begin{figure}[t]
    \centering
    \includegraphics[width=\columnwidth]{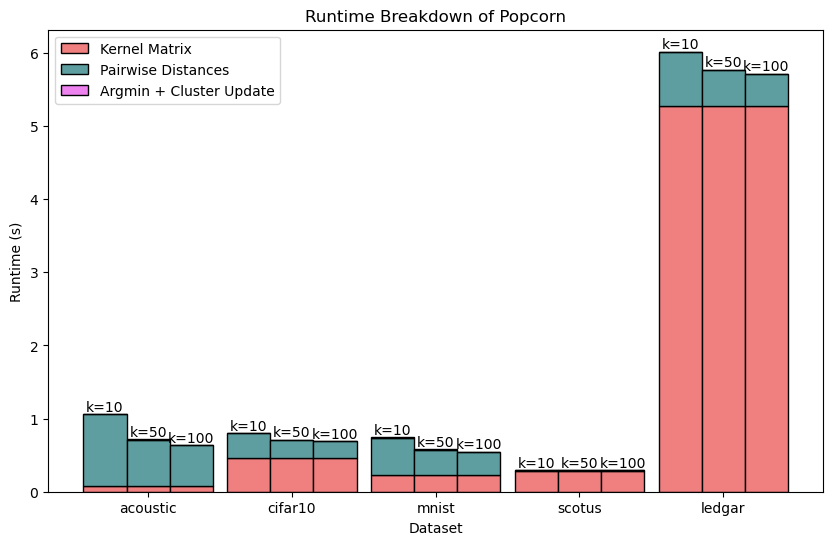}
    \caption{Runtime breakdown of \popcorn on each dataset with varying $k$. The letter dataset is excluded because it has very small runtimes.}
    \Description{}
    \label{fig:breakdown}
\end{figure}

Figure \ref{fig:breakdown} shows a runtime breakdown of \popcorn on each dataset for $k= \{10, 50, 100\}$.
The `Pairwise Distances' and `Argmin + Cluster Update' times represent sums of runtimes over 30 iterations.
Note that the letter dataset is not included in this figure, as the runtimes for that dataset are much smaller than all the others.
For datasets with large values of $d$, such as ledgar and scotus, computing the kernel matrix is more expensive than computing pairwise distances. 
This is because of the $\cO(n^2d)$ cost of computing \B, which becomes large as $d$ grows.

For datasets like acoustic and MNIST, which have large $n$ and small $d$, computing pairwise distances is more expensive than computing the kernel matrix.
This is because of the $\cO(n^2)$ work required for the SpMM each iteration.
Although the $\cO(n^2d)$ work required to compute the kernel matrix is theoretically larger than the work required for the SpMM operations across all the clustering iterations, the kernel matrix computations are done using dense matrix operations, which are better able to utilize the GPU and are therefore faster in practice for sufficiently large $n$ and small $d$. Finally, the cost of updating cluster assignments is trivial for all datasets and all values of $k$, which matches expectations.

\section{Related Work}\label{sec:related}

There is an extensive body of research focusing on developing faster \kmeans algorithms.
Elkan~\cite{elkan2003using} presents a classical \kmeans algorithm that uses the triangle inequality to avoid unnecessary pairwise distance computations, and Hamerly et al.~\cite{hamerly2015accelerating} proposes a memory-efficient version of this algorithm.
Prior works~\cite{martin2020gpukmeans, farivar2008parallel, lutz2018efficient, li2013speeding, li2023large} describe GPU implementations of classical \kmeans, but they do not emphasize the use of matrix computation.

Shawe-Taylor et al.~\cite{kernelmethods} provide a comprehensive survey of kernel methods in machine learning, including Kernel \kmeans.
Dhillon et al.~\cite{dhillon2004kernel} show that Kernel \kmeans can be used to solve spectral clustering more efficiently in cases where the data is large and sparse.
To the best of the authors' knowledge, Baydoun et al.~\cite{baydoun2018cpu} is the only previous work describing a GPU implementation of Kernel \kmeans.
This implementation is not open source, so a direct quantitative comparison is not possible.
However, Baydoun et al. \cite{baydoun2018cpu} does not explore the use of sparsity in Kernel \kmeans, and their formulation does not use SpMM or SpMV. 
It instead relies upon a formulation that uses dense matrices, which does not scale well to values of $k$ larger than 2.
Furthermore, their formulation does not use SpMV to compute centroid norms, it instead relies upon a set of hand-written reduction routines that create the need for an additional $\cO(n^2)$ work.
Finally, their work does not consider strategies for dynamically switching between different BLAS kernels when computing the kernel matrix.

The MATLAB PRMLT package \cite{chen2024prmlt} implementation of \kk uses SpMM, but it is CPU-only and does not use SpMV or dynamically switch between different strategies for computing the kernel matrix.
This paper is the first to formally describe an implementation of \kk using SpMM.

In general, most research on Kernel \kmeans has focused on developing strategies to approximate the kernel matrix ~\cite{chitta2011approximate}.
Other papers~\cite{sparsekkmeans, maldonado2015kernel} describe strategies for performing feature selection on the data before clustering it with Kernel \kmeans, yielding more accurate outcomes.
However, none of them use sparse matrix computations or GPUs.
Our use of sparse matrix computations to accelerate \kk on GPUs is orthogonal to these works and could be applied to them in future work.

Previous work has used sparse matrices to formulate and implement several parallel algorithms.
This approach is most commonly used for graph algorithms~\cite{azad2019lacc, davis2019algorithm, bulucc2011combinatorial}, but it has also found application in other areas.
For example, Solomonik et al.~\cite{solomonik2015sparse} present a general parallel programming model based on sparse tensor algebra. Guidi et al.~\cite{guidi2021parallel, guidi2022distributed, guidi2021bella} show that genomics computations can be parallelized using sparse matrix multiplication.
Tripathy et al.~\cite{tripathy2020reducing} use sparse matrices to accelerate distributed graph neural network training, and Ranawaka et al.~\cite{ranawaka2024scalable} use sparse matrix computation to scale node embedding.
Azad et al. ~\cite{azad2018hipmcl} parallelize Markov Clustering across distributed memory using sparse general matrix multiply (SpGEMM). 
Arfaoui et al. \cite{arfaoui2016efficient} use matrix multiply to implement graph algorithms on GPUs for efficient large-scale multiple-input multiple-output handling.
Gallet et al. \cite{gallet2022leveraging} implement pairwise distances on tensor cores using dense matrix operations.
Lastly, Ltaief et al. \cite{ltaief2024toward} use integer tensor cores to accelerate Euclidean distance computations for massive Genome-Wide Association Studies.

\section{Conclusion and Future Work}\label{sec:conclusion}

In this work, we presented a novel formulation of Kernel \kmeans using SpMM and SpMV and showed that this formulation enables the development of a GPU-based version of \kk with minimal manual kernel engineering effort.
In total, our implementation required less than 50 lines of handwritten CUDA code, and it is the first open-source GPU-based implementation of \kk.
Our implementation achieved orders of magnitude speedups over a comparable CPU implementation and up to $2.6\times$ over a comparable CUDA implementation on several real-world datasets.
Our results showed that a formulation based on sparse matrices is an effective tool for rapidly developing high-performance GPU-based algorithms.

In the future, we plan to explore a distributed \kk implementation based on distributed SpMM and distributed SpMV algorithms.
A distributed version of \kk would make it possible to cluster datasets that are too large to fit on a single GPU, such as multi-terabyte human activity recognition datasets \cite{jamel2020human}.

\section*{Acknowledgment}
The authors acknowledge financial support from \textit{ICSC – Centro Nazionale di Ricerca in High-Performance Computing, Big Data and Quantum Computing}, funded by European Union -- NextGenerationEU. 
This work has received funding from the European High-Performance Computing Joint Undertaking (JU) under grant agreement No 101175702 and the National Institute of Higher Mathematics Francesco Severi. 
This research used resources of the National Energy Research
Scientific Computing Center, a DOE Office of Science User Facility
supported by the Office of Science of the U.S. Department of Energy
under Contract No. DE-AC02-05CH11231 using NERSC award
ASCR-ERCAP0030076.
This project received support from the Center for Research on Programmable Plant Systems under National Science Foundation Grant No. DBI-2019674.



\bibliographystyle{unsrt}  
\bibliography{references}  

\begin{thebibliography}{10}

\bibitem{noviandy2024environmental}
Teuku~Rizky Noviandy, Irsan Hardi, Zahriah Zahriah, Rahmi Sofyan, Novi~Reandy Sasmita, Iin~Shabrina Hilal, and Ghalieb~Mutig Idroes.
\newblock Environmental and economic clustering of indonesian provinces: Insights from k-means analysis.
\newblock {\em Leuser Journal of Environmental Studies}, 2(1):41--51, 2024.

\bibitem{wielechowski2021interdependence}
Micha{\l} Wielechowski, Denys Cherevyk, Katarzyna Czech, Pavel Kotyza, {\L}ukasz Grz{\k{e}}da, and Lubos Smutka.
\newblock Interdependence between human capital determinants and economic development: K-means regional clustering approach for czechia and poland.
\newblock {\em Entrepreneurial Business and Economics Review}, 9(4):173--194, 2021.

\bibitem{melman2018k}
Paul Melman and Usman~W Roshan.
\newblock K-means-based feature learning for protein sequence classification.
\newblock {\em Proceedings of the BICOB, Las Vegas, NV, USA}, pages 19--21, 2018.

\bibitem{chen2022randomly}
Yifan Chen, Ethan~N Epperly, Joel~A Tropp, and Robert~J Webber.
\newblock Randomly pivoted cholesky: Practical approximation of a kernel matrix with few entry evaluations.
\newblock {\em arXiv preprint arXiv:2207.06503}, 2022.

\bibitem{blalock2021multiplying}
Davis Blalock and John Guttag.
\newblock Multiplying matrices without multiplying.
\newblock In {\em International Conference on Machine Learning}, pages 992--1004. PMLR, 2021.

\bibitem{dhillon2004kernel}
Inderjit~S Dhillon, Yuqiang Guan, and Brian Kulis.
\newblock Kernel k-means: spectral clustering and normalized cuts.
\newblock In {\em Proceedings of the tenth ACM SIGKDD international conference on Knowledge discovery and data mining}, pages 551--556, 2004.

\bibitem{appCancer}
Mehmet G\"{o}nen and Adam~A Margolin.
\newblock Localized data fusion for kernel k-means clustering with application to cancer biology.
\newblock In Z.~Ghahramani, M.~Welling, C.~Cortes, N.~Lawrence, and K.Q. Weinberger, editors, {\em Advances in Neural Information Processing Systems}, volume~27. Curran Associates, Inc., 2014.

\bibitem{appDiabetes}
MARTHA Alamsyah, ZUMROTUN Nafisah, E~Prayitno, AM~Afida, and EM~Imah.
\newblock The classification of diabetes mellitus using kernel k-means.
\newblock In {\em Journal of Physics: Conference Series}, volume 947, page 012003. IOP Publishing, 2018.

\bibitem{appDiabetes1}
Tru Cao, Chau Vo, Son Nguyen, Atsushi Inoue, and Duanning Zhou.
\newblock A kernel <i>k</i>-means-based method and attribute selections for diabetes diagnosis.
\newblock {\em Journal of Advanced Computational Intelligence and Intelligent Informatics}, 24(1):73--82, 2020.

\bibitem{appStudents}
Sajadin Sembiring.
\newblock {\em An Application of Predicting Student Performance Using Kernel K-Means and Smooth Support Vector Machine}.
\newblock PhD thesis, UMP, 2012.

\bibitem{appAgricolture}
Junhong Zhong and Qi~Lai.
\newblock Smart agriculture system based on internet of things using kernel k-means with support vector machine.
\newblock In {\em 2024 Second International Conference on Data Science and Information System (ICDSIS)}, pages 1--4, 2024.

\bibitem{imageChange}
Lu~Jia, Ming Li, Peng Zhang, Yan Wu, and Huahui Zhu.
\newblock Sar image change detection based on multiple kernel k-means clustering with local-neighborhood information.
\newblock {\em IEEE Geoscience and Remote Sensing Letters}, 13(6):856--860, 2016.

\bibitem{GPUdl}
Ebubekir BUBER and Banu DIRI.
\newblock Performance analysis and cpu vs gpu comparison for deep learning.
\newblock In {\em 2018 6th International Conference on Control Engineering \& Information Technology (CEIT)}, pages 1--6, 2018.

\bibitem{GPUSC}
Zhe Fan, Feng Qiu, A.~Kaufman, and S.~Yoakum-Stover.
\newblock Gpu cluster for high performance computing.
\newblock In {\em SC '04: Proceedings of the 2004 ACM/IEEE Conference on Supercomputing}, pages 47--47, 2004.

\bibitem{vestias2014trends}
Mario Vestias and Hor{\'a}cio Neto.
\newblock Trends of cpu, gpu and fpga for high-performance computing.
\newblock In {\em 2014 24th International Conference on Field Programmable Logic and Applications (FPL)}, pages 1--6. IEEE, 2014.

\bibitem{martin2020gpukmeans}
Martin Kruli\v{s} and Miroslav Kratochv\'{\i}l.
\newblock Detailed analysis and optimization of cuda k-means algorithm.
\newblock In {\em Proceedings of the 49th International Conference on Parallel Processing}, ICPP '20, New York, NY, USA, 2020. Association for Computing Machinery.

\bibitem{baydoun2018cpu}
Mohammed Baydoun, Hassan Ghaziri, and Mohammed Al-Husseini.
\newblock Cpu and gpu parallelized kernel k-means.
\newblock {\em The Journal of Supercomputing}, 74(8):3975--3998, 2018.

\bibitem{rapidsai}
Rapidsai.
\newblock Rapidsai/raft: Raft contains fundamental widely-used algorithms and primitives for data science, graph and machine learning., 2022.

\bibitem{vadim_markovtsev_2017_286944}
Vadim Markovtsev and Máximo Cuadros.
\newblock src-d/kmcuda: 6.0.0-1, 2017.

\bibitem{nickolls2008scalable}
John Nickolls, Ian Buck, Michael Garland, and Kevin Skadron.
\newblock Scalable parallel programming with cuda: Is cuda the parallel programming model that application developers have been waiting for?
\newblock {\em Queue}, 6(2):40--53, 2008.

\bibitem{Sato2010}
Katsuto Sato, Hiroyuki Takizawa, Kazuhiko Komatsu, and Hiroaki Kobayashi.
\newblock {\em Automatic Tuning of CUDA Execution Parameters for Stencil Processing}, pages 209--228.
\newblock Springer New York, New York, NY, 2010.

\bibitem{cudaHard}
Yuri Torres, Arturo Gonzalez-Escribano, and Diego~R. Llanos.
\newblock Understanding the impact of cuda tuning techniques for fermi.
\newblock In {\em 2011 International Conference on High Performance Computing \& Simulation}, pages 631--639, 2011.

\bibitem{cuSPARSE}
{NVIDIA Corporation}.
\newblock {cuSPARSE}, 2024.
\newblock \url{https://docs.nvidia.com/cuda/cusparse}.

\bibitem{cuBLAS}
{NVIDIA Corporation}.
\newblock {cuBLAS}, 2024.
\newblock \url{https://docs.nvidia.com/cuda/cublas}.

\bibitem{lloyd}
S.~Lloyd.
\newblock Least squares quantization in pcm.
\newblock {\em IEEE Transactions on Information Theory}, 28(2):129--137, 1982.

\bibitem{kmeans++}
David Arthur and Sergei Vassilvitskii.
\newblock k-means++: the advantages of careful seeding.
\newblock In {\em Proceedings of the Eighteenth Annual ACM-SIAM Symposium on Discrete Algorithms}, SODA '07, page 1027–1035, USA, 2007. Society for Industrial and Applied Mathematics.

\bibitem{kernelmethods}
John Shawe-Taylor and Nello Cristianini.
\newblock {\em Kernel methods for pattern analysis}.
\newblock Cambridge university press, 2004.

\bibitem{Bell2012ThrustAP}
Nathan Bell and Jared Hoberock.
\newblock Thrust: A productivity-oriented library for cuda, 2012.

\bibitem{chang2011libsvm}
Chih-Chung Chang and Chih-Jen Lin.
\newblock Libsvm: A library for support vector machines.
\newblock {\em ACM transactions on intelligent systems and technology (TIST)}, 2(3):1--27, 2011.

\bibitem{6737057}
Karl Ni and Ryan Prenger.
\newblock Learning features in deep architectures with unsupervised kernel k-means.
\newblock In {\em 2013 IEEE Global Conference on Signal and Information Processing}, pages 981--984, 2013.

\bibitem{large}
Rong Zhang and A.I. Rudnicky.
\newblock A large scale clustering scheme for kernel k-means.
\newblock In {\em 2002 International Conference on Pattern Recognition}, volume~4, pages 289--292 vol.4, 2002.

\bibitem{chen2024prmlt}
Mo~Chen.
\newblock Pattern recognition and machine learning toolbox.
\newblock \url{https://github.com/PRML/PRMLT}, 2024.
\newblock Retrieved July 30, 2024.

\bibitem{roofline}
Samuel Williams, Andrew Waterman, and David Patterson.
\newblock Roofline: an insightful visual performance model for multicore architectures.
\newblock {\em Commun. ACM}, 52(4):65–76, apr 2009.

\bibitem{ncu}
{NVIDIA Corporation}.
\newblock {NVIDIA Nsight Compute}, 2024.
\newblock \url{https://docs.nvidia.com/nsight-compute/index.html}.

\bibitem{elkan2003using}
Charles Elkan.
\newblock Using the triangle inequality to accelerate k-means.
\newblock In {\em Proceedings of the 20th international conference on Machine Learning (ICML-03)}, pages 147--153, 2003.

\bibitem{hamerly2015accelerating}
Greg Hamerly and Jonathan Drake.
\newblock Accelerating lloyd’s algorithm for k-means clustering.
\newblock {\em Partitional clustering algorithms}, pages 41--78, 2015.

\bibitem{farivar2008parallel}
Reza Farivar, Daniel Rebolledo, Ellick Chan, and Roy~H Campbell.
\newblock A parallel implementation of k-means clustering on gpus.
\newblock In {\em Pdpta}, volume~13, pages 212--312, 2008.

\bibitem{lutz2018efficient}
Clemens Lutz, Sebastian Bre{\ss}, Tilmann Rabl, Steffen Zeuch, and Volker Markl.
\newblock Efficient k-means on gpus.
\newblock In {\em Proceedings of the 14th International Workshop on Data Management on New Hardware}, pages 1--3, 2018.

\bibitem{li2013speeding}
You Li, Kaiyong Zhao, Xiaowen Chu, and Jiming Liu.
\newblock Speeding up k-means algorithm by gpus.
\newblock {\em Journal of Computer and System Sciences}, 79(2):216--229, 2013.

\bibitem{li2023large}
Mi~Li, Eibe Frank, and Bernhard Pfahringer.
\newblock Large scale k-means clustering using gpus.
\newblock {\em Data Mining and Knowledge Discovery}, 37(1):67--109, 2023.

\bibitem{chitta2011approximate}
Radha Chitta, Rong Jin, Timothy~C Havens, and Anil~K Jain.
\newblock Approximate kernel k-means: Solution to large scale kernel clustering.
\newblock In {\em Proceedings of the 17th ACM SIGKDD international conference on Knowledge discovery and data mining}, pages 895--903, 2011.

\bibitem{sparsekkmeans}
Xin Guan and Yoshikazu Terada.
\newblock Sparse kernel k-means for high-dimensional data.
\newblock {\em Pattern Recognition}, 144:109873, 2023.

\bibitem{maldonado2015kernel}
Sebasti{\'a}n Maldonado, Emilio Carrizosa, and Richard Weber.
\newblock Kernel penalized k-means: A feature selection method based on kernel k-means.
\newblock {\em Information sciences}, 322:150--160, 2015.

\bibitem{azad2019lacc}
Ariful Azad and Ayd{\i}n Bulu{\c{c}}.
\newblock Lacc: A linear-algebraic algorithm for finding connected components in distributed memory.
\newblock In {\em 2019 IEEE International Parallel and Distributed Processing Symposium (IPDPS)}, pages 2--12. IEEE, 2019.

\bibitem{davis2019algorithm}
Timothy~A Davis.
\newblock Algorithm 1000: Suitesparse: Graphblas: Graph algorithms in the language of sparse linear algebra.
\newblock {\em ACM Transactions on Mathematical Software (TOMS)}, 45(4):1--25, 2019.

\bibitem{bulucc2011combinatorial}
Ayd{\i}n Bulu{\c{c}} and John~R Gilbert.
\newblock The combinatorial blas: Design, implementation, and applications.
\newblock {\em The International Journal of High Performance Computing Applications}, 25(4):496--509, 2011.

\bibitem{solomonik2015sparse}
Edgar Solomonik and Torsten Hoefler.
\newblock Sparse tensor algebra as a parallel programming model.
\newblock {\em arXiv preprint arXiv:1512.00066}, 2015.

\bibitem{guidi2021parallel}
Giulia Guidi, Oguz Selvitopi, Marquita Ellis, Leonid Oliker, Katherine Yelick, and Ayd{\i}n Bulu{\c{c}}.
\newblock Parallel string graph construction and transitive reduction for de novo genome assembly.
\newblock In {\em 2021 IEEE International Parallel and Distributed Processing Symposium (IPDPS)}, pages 517--526. IEEE, 2021.

\bibitem{guidi2022distributed}
Giulia Guidi, Gabriel Raulet, Daniel Rokhsar, Leonid Oliker, Katherine Yelick, and Aydin Buluc.
\newblock Distributed-memory parallel contig generation for de novo long-read genome assembly.
\newblock In {\em Proceedings of the 51st International Conference on Parallel Processing}, pages 1--11, 2022.

\bibitem{guidi2021bella}
Giulia Guidi, Marquita Ellis, Daniel Rokhsar, Katherine Yelick, and Ayd{\i}n Bulu{\c{c}}.
\newblock Bella: Berkeley efficient long-read to long-read aligner and overlapper.
\newblock In {\em SIAM Conference on Applied and Computational Discrete Algorithms (ACDA21)}, pages 123--134. SIAM, 2021.

\bibitem{tripathy2020reducing}
Alok Tripathy, Katherine Yelick, and Ayd{\i}n Bulu{\c{c}}.
\newblock Reducing communication in graph neural network training.
\newblock In {\em SC20: International Conference for High Performance Computing, Networking, Storage and Analysis}, pages 1--14. IEEE, 2020.

\bibitem{ranawaka2024scalable}
Isuru Ranawaka and Ariful Azad.
\newblock Scalable node embedding algorithms using distributed sparse matrix operations.
\newblock In {\em 2024 IEEE International Parallel and Distributed Processing Symposium Workshops (IPDPSW)}, pages 1199--1201. IEEE, 2024.

\bibitem{azad2018hipmcl}
Ariful Azad, Georgios~A Pavlopoulos, Christos~A Ouzounis, Nikos~C Kyrpides, and Aydin Bulu{\c{c}}.
\newblock Hipmcl: a high-performance parallel implementation of the markov clustering algorithm for large-scale networks.
\newblock {\em Nucleic acids research}, 46(6):e33--e33, 2018.

\bibitem{arfaoui2016efficient}
Mohamed-Amine Arfaoui, Hatem Ltaief, Zouheir Rezki, Mohamed-Slim Alouini, and David Keyes.
\newblock Efficient sphere detector algorithm for massive mimo using gpu hardware accelerator.
\newblock {\em Procedia Computer Science}, 80:2169--2180, 2016.

\bibitem{gallet2022leveraging}
Benoit Gallet and Michael Gowanlock.
\newblock Leveraging gpu tensor cores for double precision euclidean distance calculations.
\newblock In {\em 2022 IEEE 29th International Conference on High Performance Computing, Data, and Analytics (HiPC)}, pages 135--144. IEEE, 2022.

\bibitem{ltaief2024toward}
Hatem Ltaief, Rabab Alomairy, Qinglei Cao, Jie Ren, Lotfi Slim, Thorsten Kurth, Benedikt Dorschner, Salim Bougouffa, Rached Abdelkhalak, and David~E Keyes.
\newblock Toward capturing genetic epistasis from multivariate genome-wide association studies using mixed-precision kernel ridge regression.
\newblock {\em arXiv preprint arXiv:2409.01712}, 2024.

\bibitem{jamel2020human}
Ahmed~AM Jamel and Bahriye Akay.
\newblock Human activity recognition based on parallel approximation kernel k-means algorithm.
\newblock {\em Computer Systems Science \& Engineering}, 35(6), 2020.

\end{thebibliography}

\begin{appendices}

\section{Artifact Description}

\subsection{Availability}
Our artifact is available at Zenodo \url{https://doi.org/10.5281/zenodo.14219596}.
This archive contains the code for our artifact, scripts for reproducing results in the paper, and a README largely identical to this one. 
\subsection{Requirements}
\subsubsection{Hardware Requirements}
\begin{itemize}
    \item Modern x86-64 CPU.
    \item One NVIDIA A100 GPU with 80GB of HBM
\end{itemize}
\subsubsection{Software Requirements}
\begin{itemize}
    \item cmake >=3.24.3
    \item cudatoolkit 12.2 
    \item GCC 12.3
    \item RAFT 24.08.00. cmake will take care of installing this for you.
    \item python >=3.10
    \item MATLAB R2023b or later
\end{itemize}
\subsection{Getting Started}

In order to obtain our artifact, please \texttt{wget \url{https://zenodo.org/records/14219596/files/ppopp25_popcorn.tar.gz}}, then decompress the tarball and \texttt{cd} into the \texttt{popcorn\_artifact} directory.
From here, run
\begin{lstlisting}
mkdir build && cd build
cmake ..
\end{lstlisting}
A minor fix to the \texttt{fmt} library is necessary in order to allow RAFT to compile. 

Open \texttt{build/\_deps/fmt-src/include/fmt/format-inl.h} and comment out lines 147-150 and lines 1407-1408.
This will allow RAFT to compile successfully. 
Please note that this minor patch will not impact the performance or correctness of any implementation evaluated in our experiments, since \texttt{fmt} is only used to format output, and the lines that are commented out are not executed by Popcorn or any baseline implementation.

Once these changes have been made, run
\begin{lstlisting}
    make -j
\end{lstlisting}

It will take some time to compile RAFT and its dependencies, so the whole build process is expected to take around 10-15 minutes. Any compiler warnings can be safely ignored.
This will generate \texttt{src/bin/gpukmeans}, which can be used to run Popcorn.

\subsection{Running Popcorn}
To run \popcorn independently, one should use the following command line arguments:

\begin{itemize}
    \item \verb|-n INT|: Number of data points in dataset
    \item \verb|-d INT|: Dimensionality of dataset
    \item \verb|-k INT|: Number of clusters
    \item \verb|--runs INT|: Number of times to run the clustering
    \item \verb|-t FLOAT|: Tolerance for determining convergence
    \item \verb|-m INT|: Maximum number of iterations to run clustering for
    \item -c \{0|1\} : Whether or not to check for convergence. \verb|0| will run \popcorn for however many iterations are specified in the \verb|-m| argument. \verb|1| will run \popcorn until convergence or until the maximum number of iterations has been executed.
    \item \verb|--init STR|: Method for centroid initialization. Should always be set to \verb|random|.
    \item \verb|-f STR|: Kernel function. Should be one of \verb|linear|, \verb|polynomial|, or \verb|sigmoid|.
    \item \verb|-i STR|: Path to input file. Should be stored in libsvm format or as a standard CSV. If not set, a random dataset is initialized.
    \item \verb|-s INT|: Seed to use for RNG.
    \item -l \{0|2\} Select GPU implementation. \verb|0| runs the naive baseline, \verb|2| runs \popcorn.
    \item \verb|-o STR|: Write clustering results to a file.
\end{itemize}

\subsection{Reproducing Results}
To run experiments, cd into the experiments directory and run \texttt{run.sh}. This script will also generate plots from the csv files. The first set of experiments that will run benchmark Popcorn. The next set benchmarks the naive GPU implementation. The last set benchmarks the MATLAB CPU implementation.

It should take no more than 3 hours to run all experiments, since the CPU implementation is quite slow for the larger datasets.

3 plots should be produced by the script; 'speedup-cuda.png', 'distances-speedup.png', and 'speedup-popcorn.'png'. 'speedup-cuda.png' shows the speedup of the baseline GPU implementation over the CPU implementation (Fig 4 in the paper). 'distances-speedup.png' shows the speedup of Popcorn over the GPU baseline, only considering the time it takes to peform the clustering (Fig 5 in the paper). Lastly, 'speedup-popcorn.png' shows the speedup of the entirety of Popcorn's kernel k-means algorithm over the entirety of the baseline GPU implementation's kernel k-means algorithm (Fig 8 in the paper).

\end{appendices}

\end{document}